%% file: main.tex
\documentclass{article}

\usepackage{arxiv}

\usepackage[utf8]{inputenc} 
\usepackage[T1]{fontenc}    
\usepackage{hyperref}       
\usepackage{url}            
\usepackage{booktabs}       
\usepackage{amsfonts}       
\usepackage{nicefrac}       
\usepackage{microtype}      
\usepackage{graphicx}
\usepackage{booktabs}
\usepackage{csquotes}
\usepackage{mdframed}
\usepackage{array}
\usepackage{subcaption}
\usepackage{dblfloatfix}
\usepackage[section]{placeins}
\usepackage{enumitem}
\usepackage{amsmath}
\usepackage{float}
\usepackage{wrapfig}
\usepackage{caption}

\title{Models Matter: \\Setting Accurate Privacy Expectations for Local and Central Differential Privacy}

\author{
    Mary Anne Smart\thanks{This work was completed while Mary Anne Smart was at UC San Diego.}\\
    masmart@purdue.edu\\
    Purdue University\\
\And
    Priyanka Nanayakkara\\
    priyankan@u.northwestern.edu\\
    Northwestern University\\
\And
    Rachel Cummings\\
    rac2239@columbia.edu\\
    Columbia University\\
\And
    Gabriel Kaptchuk\thanks{This work was completed while Gabriel Kaptchuk was at Boston University.}\\
    kaptchuk@umd.edu\\
    University of Maryland, College Park\\
\And
    Elissa Redmiles\\
    eredmiles@gmail.com\\
    Georgetown University\\
}

\input{macros}
\begin{document}
\maketitle
\begin{abstract}
\input{0_abstract}
\end{abstract}

\section{Introduction}
\input{1_introduction}
\section{Background}
\input{2_background}
\section{Interview Study}
\input{3_interviews}
\section{Large-Scale Evaluation}
\label{sec:eval}
\input{4_survey}
\section{Limitations}
\input{5_limitations}
\section{Discussion}
\input{6_discussion}

\section*{Acknowledgments}
We would like to thank everyone who provided feedback on various stages of this project, especially Aaron Broukhim, participants in the Technically Private reading group. All authors were supported by DARPA (contract number W911NF-21-1-0371). Any opinions, findings, and conclusions or recommendations expressed in this material are those of the authors and do not necessarily reflect the views of the United States Government or DARPA. In addition to DARPA support, the third author was supported in part by NSF grant CNS-1942772 (CAREER), a Mozilla Research Grant, a JP-Morgan Chase Faculty Research Award, and an Apple Privacy-Preserving Machine Learning Award. The fourth author was also supported by NSF grant \#2030859 to the Computing Research Association for the CIFellows Project, and the fifth author was also supported by a Google Research Scholar Award.
\newpage

\bibliographystyle{acm}
\bibliography{references,abbrev3,crypto}

\appendix
\input{7_appendix}

\end{document}

%% file: macros.tex
\definecolor{Orange}{rgb}{1,0.5,0}
\definecolor{Red}{rgb}{1,0,0}
\definecolor{Green}{RGB}{0,150,0}
\definecolor{Purple}{rgb}{0.75,0,1}
\definecolor{babypink}{rgb}{0.96, 0.76, 0.76}
\definecolor{azure}{rgb}{0,0.49,1}
\definecolor{periwinkle}{rgb}{0.8, 0.8, 1.0}
\definecolor{Pink}{RGB}{204, 0, 204}
\definecolor{DarkGreen}{RGB}{20, 89, 38}
\definecolor{DarkRed}{RGB}{99, 12, 9}

\newlength{\cmfapaparindent}
\SetBlockEnvironment{myquote} 

\ifdefined\commentsenabled
\newcommand{\elissa}[1]{\textsf{\textbf{\textcolor{Purple}{[[EMR: #1]]}}}}
\newcommand{\gsk}[1]{\textsf{\textbf{\textcolor{azure}{[[GSK: #1]]}}}}
\newcommand{\priyanka}[1]{\textcolor{red}{PN: #1}}
\newcommand{\mas}[1]{\textcolor{Green}{MAS: #1}}
\newcommand{\rc}[1]{\textcolor{Orange}{RC: #1}}
\else
\newcommand{\elissa}[1]{}
\newcommand{\gsk}[1]{}
\newcommand{\priyanka}[1]{}
\newcommand{\mas}[1]{}
\newcommand{\rc}[1]{}
\fi

\urldef\myurl\url{https://osf.io/3acvw/?view_only=f12174861ffd4cd0872a54a8e1326a26}
\urldef\nounurl\url{https://thenounproject.com}

\newcommand{\linReg}[2]{$\beta=#1$; $p#2$}
\newcommand{\ordReg}[2]{$\text{OR}=#1$; $p#2$}

\renewcommand{\paragraph}[1]{\smallskip \noindent \textbf{#1}}

%% file: 0_abstract.tex
Differential privacy is a popular privacy-enhancing technology that has been deployed both in industry and government agencies. Unfortunately, existing explanations of differential privacy fail to set accurate privacy expectations for data subjects, which depend on the choice of deployment model. We design and evaluate new explanations of differential privacy for the local and central models, drawing inspiration from prior work explaining other privacy-enhancing technologies. We find that consequences-focused explanations in the style of privacy nutrition labels that lay out the implications of differential privacy are a promising approach for setting accurate privacy expectations. Further, we find that while process-focused explanations are not enough to set accurate privacy expectations, combining consequences-focused explanations with a brief description of how differential privacy works leads to greater trust. 

%% file: 1_introduction.tex
Usable privacy research has long focused on making information about privacy protections transparent to end users in order to allow for informed decision-making about data sharing~\cite{lipford2009visible,felt2016rethinking,emami2022informative}.
Prior work has sought to explain privacy enhancing technologies (PETs) such as end-to-end encryption in messaging~\cite{demjaha2018metaphors} and HTTPS/TLS \cite{10.1145/2556288.2557292}. 
Existing explanation strategies vary in terms of how much they try to explain data protection mechanisms or processes versus implications of the particular PET.
As PETs become increasingly complex and offer more nuanced notions of privacy, there is a significant need for increased research into best practice for transparent PET messaging that matches these new techniques and empowers end users.

Differential privacy (DP) \cite{TCC:DMNS06} is a privacy-enhancing technology that has quickly been adopted by industry and government agencies~\cite{erlingsson2014rappor, fb2020, thakurta2017learning, abowd2018us, tumult,desfontainesblog20211001}.
DP deployments provide provable privacy guarantees by adding statistical noise to data or computations; this noise obfuscates the information of each individual while preserving aggregate-level insights. In response to DP's rapid success, a growing body of work has started to document the inadequacies of existing messaging around DP~\cite{CCS:CumKapRed21} and design new messaging techniques for DP systems~\cite{xiong2020towards,Xiong2022UsingIT,karegar2022exploring,nanayakkara2023chances,Bullek2017,smart2022,CCS:FVSTM22}. 

The technical and mathematical complexity of PETs like DP makes effective communication challenging~\cite{CCS:CumKapRed21, nanayakkara2022s, abu2018exploring,demjaha2018metaphors}. DP is an interesting case study for constructing transparent PET messaging because it is an instance of an emerging PET paradigm that has received relatively little attention: privacy-preserving, outsourced computation. This paradigm is increasingly important as more users rely on computationally-weak mobile devices~\cite{pew_mobile}. In response to this growing need, new approaches to privacy-preserving computational outsourcing are being actively developed, both in industry~\cite{awsnitro,sealfhe} and academia~\cite{10.1145/3209811.3212701}. 
Although there has been some work on people's mental models of other PETs in this category~\cite{kacsmar2022comprehension} and creating transparent messaging for functional encryption in particular~\cite{alaqra2023structural}, further work is needed. DP is particularly interesting because of its widespread deployment and use of statistical methods (as opposed to encryption). 

In this work, we develop messaging for DP that highlights the threat models that are implicit in different approaches to deploying DP.  These implicit threat models are critical for end users to understand before sharing their data, as the chosen threat model may not provide protections against the classes of attackers about which they are concerned.  
We do this by exploring three explanation formats drawn from the existing PETs messaging literature: nutrition labels~\cite{kelley2009nutrition}, diagrams~\cite{Xiong2022UsingIT,smart2022,stransky2021limited}, and metaphors~\cite{ZhangKennedy,demjaha2018metaphors,raja2011brick,karegar2022exploring,zhang2016role,suh2022privacytoon}. Each of our evaluated explanations aims to communicate the consequences of DP in terms of which information flows the deployment protects against. Prior work has tended to focus on explaining individual PETs in isolation, and while we focus on DP, we discuss how our designs may be modified to account for the effects of multiple PETs deployed together. 

\smallskip
\noindent
\textbf{Differential Privacy Deployment Models.}
There are multiple deployment \emph{models} for DP, each of which is associated with a particular threat model.
The two most widely deployed models are the \emph{central model}~\cite{dwork2006calibrating} and the \emph{local model}~\cite{kasiviswanathan2011can}.\footnote{Although there are many variations of DP~\cite{desfontaines2020sok}, we focus on 
the \emph{central} and \emph{local} models due to their popularity.} 

\looseness-1 The central model assumes there exists a data curator who is \emph{trusted} to see raw data from individuals; the adversary can only access released, aggregate results. The data curator collects data from individuals, performs statistical analyses on the collected dataset, and then injects statistical noise into the results before release.  This process limits the ability of the adversary to reconstruct individual records from summary statistics, at the cost of reduced accuracy. The potential danger of this model, however, is that 
the data curator might \textit{not} be trustworthy; the database storing individuals' data could be vulnerable to hackers or misuse by insiders if other, complementary security practices are not adopted in tandem. Well-known deployments of the central model include the U.S. Census Bureau's data products for the 2020 Decennial Census~\cite{abowd2018us}.

In the local model, noise is added to each individual's data \emph{before} collection, meaning the unmodified data are never stored together. As such, there is not the same need to trust the data curator (i.e., it is assumed that the data curator is honest but curious). This higher level of security comes at a cost: significantly more noise must be added to the data in order to ensure the same level of privacy protection, reducing the accuracy---and, thus, utility---of the collected data. Notably, Google and Apple have both used local DP to analyze browser data in Chrome and Safari, respectively~\cite{erlingsson2014rappor,apple2017learning}. 

\smallskip
\noindent
\textbf{Helping Users Understand DP Models.}
Ensuring that descriptions of DP accurately convey information about the model is crucial to designing transparent messaging for DP deployments: the threat surface associated with the two main models differ significantly, even if they provide the same privacy guarantees for the \emph{data releases}. Specifically, data collected under the central model can be hacked, leaked, or abused by an insider threat if sufficient complementary privacy measures are not taken, while data collected under the local model does not share these risks. 

Data subjects cannot be expected to make informed data-sharing decisions if they believe that DP is ``some sort of crypto-magic to protect people from data misuse''~\cite{rogaway2015moral}.
Prior work has demonstrated that existing DP description strategies do a poor job aligning users' privacy expectations with the privacy protections provided by DP models~\cite{CCS:CumKapRed21}. In other words, the kind of protection that users expect does not align with the actual nature of the protection offered by DP.
Misaligned expectations exacerbated by poor communication can lead to data subjects underestimating or---even more alarmingly---overestimating the privacy protection that DP offers. 

The goal of our work is to find effective ways to incorporate information about the model into descriptions of DP deployments. More specifically, we use a mixed-methods approach to study the following research question:
\begin{quote}
\textbf{What are effective design strategies for explanations that help people understand which information flows are protected by DP, given a deployment model?}     
\end{quote}

We first explore three kinds of explanations that build on prior work ~\cite{karegar2022exploring,nist,Xiong2022UsingIT,kelley2009nutrition,raja2011brick,CCS:CumKapRed21} through an interview study: metaphors, diagrams, and privacy labels for DP. 
Based on the results of this study, we identify the most promising strategies---privacy labels and metaphors---and further refine these explanations based on participant feedback. 
We evaluate our refined explanations in an online survey ($n=698$), measuring objective comprehension, subjective understanding, perceived thoroughness, and trust. We compare our explanations against existing state-of-the-art text-based explanations of DP~\cite{xiong2020towards}. Based on our results, we make suggestions for future research and design of explanations of PETs. 

The process of investigating design strategies also allowed us to characterize mental models people form around DP. While studying mental models was not the primary goal of our study, we include particularly interesting insights that can guide future work. For example, we found that participants often tried to make sense of DP through comparisons to other PETs such as encryption. We conclude with a discussion of the potential design implications of our findings. 

%% file: 2_background.tex
A growing body of work provides guidance for effective S\&P communication~\cite{schaub-15-effective-notices,gluck2016short}. Awkward interfaces or ineffective communication can lead to dangerous misconceptions and risky behaviors~\cite{whitten1999johnny, krombholz2019if, CCS:CumKapRed21,habib2018away}. One reason that people may misjudge privacy risks or misuse PETs is that they lack appropriate \emph{mental models}. Prior work has argued that ``efficacy of risk communication depends not only on the nature of the risk, but also on the alignment between the conceptual model embedded in the risk communication and the user's mental model of the risk''~\cite{asgharpour2007mental}. Unfortunately, existing depictions of DP appear to be misaligned with people's mental models, resulting in misaligned privacy expectations~\cite{CCS:CumKapRed21}.

In this section, we outline the relevant prior work on the challenges of designing effective, transparent communication about PETs. First, we discuss the prior work on communicating with data subjects about DP. Next, we discuss three particularly popular privacy explanation strategies---metaphors, diagrams, and nutrition labels. Finally, we discuss prior work on mental models in S\&P. 

\input{tables/disclosures}

\smallskip
\noindent
\textbf{Implications vs Process.}
One of the most important findings from prior work is that explaining data protection processes is not enough for most readers to grasp the implications of the protection offered by PETs~\cite{smart2022, CCS:CumKapRed21,xiong2020towards,distler2020making}. Xiong et al.~\cite{xiong2020towards} studied explanations of both central and local DP, and found that when the implications of the local and central models were stated explicitly, participants were more willing to share information under the local model. Kühtreiber et al.~\cite{kuhtreiber2022replication} replicated this study with German participants. Differently from these studies, we explore a variety of best-practice methods from the usable S\&P literature (i.e., metaphors, diagrams, and nutrition labels) to communicate which information flows are protected by DP under the central and local models. 

Cummings et al.~\cite{CCS:CumKapRed21} explored the implications of DP through six information disclosures about which people care and DP may protect---depending on whether the local or central model is used. They found that existing descriptions of DP fail to appropriately set privacy expectations regarding these disclosures, in part because many descriptions are not specific to the model (e.g., central or local) being used. In contrast to this work, we build new explanations rather than evaluating existing ones. We draw on their framework to present the implications of DP (entities to whom data can potentially be disclosed) as part of our nutrition labels. For improved clarity, in our study, we combine two of the disclosures (organization and data analyst), resulting in five total (Table~\ref{tab:disclosures}). 

\looseness-1 Finally, Frazen et al.~\cite{CCS:FVSTM22} and Nanayakkara et al.~\cite{nanayakkara2023chances} developed methods of explaining the implications of the privacy budget, drawing from the risk communication literature. Nanayakkara et al.~\cite{nanayakkara2023chances} found that participants were more willing to share information as the privacy budget decreased (i.e., protections were strengthened). In our study, we assume a small privacy budget (i.e., strong privacy), so that we can focus on the implications of the deployment model. Future work could consider combining our explanations with explanations of the privacy budget. 

\smallskip
\noindent
\textbf{Metaphors}
\label{sec:background:metaphors}
\looseness-1 Metaphors are one approach for improving mental models, and have been studied extensively in the S\&P domain~\cite{ZhangKennedy,demjaha2018metaphors,raja2011brick,karegar2022exploring,zhang2016role,suh2022privacytoon}. For example, physical security metaphors can improve users' understanding of personal firewalls~\cite{raja2011brick}. In other cases, however, metaphors have been less effective. For example, descriptions of end-to-end encryption using metaphors failed to improve understanding~\cite{demjaha2018metaphors}. Prior work has also begun to explore the effectiveness of metaphors specifically for explaining DP~\cite{karegar2022exploring}. They find that functional metaphors can be useful for explaining both that injected randomness protects privacy and that there exists a tradeoff between privacy and accuracy. The metaphors we develop are also functional (i.e., focused on \textit{what} DP offers), rather than structural (i.e. focused on \textit{how} DP works)~\cite{alaqracommunicating}. While the metaphors from prior work aim to cover a long list of facts about DP, they are not designed to emphasize the different kinds of disclosures against which DP may or may not protect---the focus of our work. 

\smallskip
\noindent
\textbf{Diagrams}
\label{sec:background:diagrams}
Another strategy for explaining PETs is the use of visualizations. For example, hypothetical outcome plots~\cite{hullman2015hypothetical} have been used to visualize the protection offered by DP~\cite{smart2022, pair}; they have also been used to visualize DP's accuracy implications for data curators~\cite{nanayakkara2022visualizing}. 
In the case of randomized response~\cite{Warner1965}---a simple instantiation of local DP---the injected noise can be represented through a spinner~\cite{Bullek2017,DCHL23}. 
Recent work has also explored the use of diagrams and animations in the specific context of location privacy~\cite{Xiong2022UsingIT}.
Diagrams have also been used to explain other PETs such as encryption~\cite{stransky2021limited}. We build on this prior work to develop diagrams for DP in both the local and central models.

\smallskip
\noindent
\textbf{Nutrition Labels}
\label{sec:background:labels}
One influential approach in privacy communication broadly has been the use of ``nutrition labels'' for privacy~\cite{kelley2009nutrition}. Drawing inspiration from standardized nutrition labels on food products, privacy labels have been proposed as an alternative or supplement to typical privacy policies with their notorious usability issues~\cite{Obar2020, Turow2018}. Privacy labels have proven to be a useful way to present privacy-related information~\cite{kelley2009nutrition}. Organizing key information into carefully-designed labels helps users find information more quickly than they would by perusing a traditional privacy policy~\cite{kelley2010standardizing}. Although originally proposed for websites, similar labels have since been developed for datasets~\cite{holland2020dataset} and Internet of Things devices~\cite{emami2022informative}. Nutrition labels have even been proposed for describing DP~\cite{xiong2020effect,CCS:CumKapRed21}. 
Apple has recently integrated the nutrition labels approach into their iOS app ecosystem. Unfortunately, the utility of these labels has been hampered by the fact that labels are not always easy to find and can be misleading or inaccurate~\cite{cranor2022mobile,kollnig2022goodbye}. Our work adapts the privacy label concept for the purpose of explaining DP---specifically for explaining how local and central DP may or may not protect against particular disclosure risks. \looseness=-1

\paragraph{Mental Models.}
Mental models refer to simplified versions of complex processes that people mentally hold and which may help them understand key pieces of information~\cite{craik1967nature, lin2012expectation}. Researchers have argued that mental models are important for effectively communicating security risks to end users~\cite{stewart2012death}. Camp~\cite{camp2009mental} argues that a medical or public health mental model is particularly useful for conveying the implications of malicious code---in particular, ``that everyone is at risk,'' ``the importance and continued autonomy in the face of risk'' and the ''shared responsibility for community health.'' In this way, mental models can rely on people's existing knowledge to help them better grasp attributes of a new setting. 

However, flawed mental models can lead to dangerous decisions~\cite{vaniea2014mental, wash2010folk}. For example, Wash~\cite{wash2010folk} proposes folk models of viruses and hackers and describes how these models help explain why people ignore security advice. A mental models approach can also clarify how people's backgrounds may impact their understanding of risks~\cite{kang2015my, oates2018turtles, bravo-lillo-11-warning-mental}. For instance, people's level of computer science background impacts the complexity of their internet mental models, and therefore the number of privacy threats they perceive~\cite{kang2015my}. Oates et al.~\cite{oates2018turtles} find that when asked to create illustrations of the meaning of privacy, experts' illustrations tend to depict privacy as more ``nuanced'' than non-experts' illustrations. Bravo-Lillo et al.~\cite{bravo-lillo-11-warning-mental} also find that novice and advanced users have different mental models and risk perceptions.

Finally, researchers have noted the value of studying privacy expectations~\cite{lin2012expectation, rao-16-unexpected}. For example, Lin et al.~\cite{lin2012expectation} propose evaluating mobile app privacy by studying people's privacy expectations of apps, while Rao et al.~\cite{rao-16-unexpected} suggest that understanding misalignment's between people's expectations and privacy policies can help reduce privacy risks.

%% file: tables/disclosures.tex
\begin{table}[tbh]
  \centering
  \caption{ Five Information Disclosures. We combine the ``organization'' and ``data analyst'' categories from prior work, since a data analyst is simply an employee of the organization~\cite{CCS:CumKapRed21}. Although some implementations of the central model limit employees' access to the data (e.g., Uber~\cite{johnson2020chorus}), we consider the more common case where only published information is privacy-protected.}
  \begin{tabular}{p{13cm} c c}
    \toprule
    \textbf{Information Disclosure}
    & \textbf{Local} & \textbf{Central}\\
    \midrule
    {\small 
    \textbf{Hack:} \textit{A criminal or foreign government that hacks the non-profit could learn my medical history.}
    } 
    & 
    {\small 
    False
    } 
    & 
    {\small 
    True
    } 
    \\
    {\small 
    \textbf{Law:} \textit{A law enforcement organization could access my medical history with a court order requesting this data from the non-profit.}
    } 
    & 
    {\small 
    False
    } 
    & 
    {\small 
    True
    } 
    \\
    {\small 
    \textbf{Org:} \textit{An employee working for the non-profit, such as a data analyst, could be able to see my exact medical history.}
    } 
    & 
    {\small
    False 
    }
    & 
    {\small
    True 
    }
    \\
    {\small 
    \textbf{Graph:} \textit{Graphs or informational charts created using information given to the non-profit could reveal my medical history.}
    } 
    & 
    {\small 
    False
    } 
    & 
    {\small 
    False
    } 
    \\
    {\small 
    \textbf{Share:} \textit{Data that the non-profit shares with other organizations doing medical research could reveal my medical history.}
    } 
    & 
    {\small 
    False
    } 
    & 
    {\small 
    True
    } 
    \\
    \bottomrule
  \end{tabular}
  ~\label{tab:disclosures}
\end{table}

%% file: 3_interviews.tex
We began designing explanations by developing a set of initial prototypes, drawing from prior work in S\&P communication. Through an interview study,\footnote{All studies were approved by the Human Research Protection Office.} we use these prototypes to solicit feedback on what makes an effective explanation of DP. 

\paragraph{Scenario}
We situate our designs within the medical data collection scenario from~\cite{CCS:CumKapRed21}. In this scenario, a non-profit organization is collecting health data for medical research. Because medical information is considered highly sensitive~\cite{ion2011home, schomakers2019internet}, data subjects are more likely to care about understanding the privacy implications of DP in this scenario. 

\subsection{Initial Prototypes}
Metaphors can help non-experts develop more useful mental models. We develop four initial metaphors---two for the local model and two for the central model---designed to clarify the kinds of risk involved. 
All four can be found in Appendix~\ref{app:designs}.

We also draw inspiration from prior work on visualizations of DP~\cite{Bullek2017,nanayakkara2022visualizing,pair,smart2022,nist,nanayakkara2023chances} to design our own diagrams that highlight how DP protects or fails to protect against the disclosures listed in Table~\ref{tab:disclosures}. We developed our diagrams through an iterative process. We discussed the accuracy and clarity of initial diagrams as a group, and based on the discussion, iterated on our designs.  In the end, we developed four diagrams---two for the local model and two for the central model---with slight differences in iconography. After initial interviews, we added a third variation for both models that included a caption. All diagrams used a vertical line to depict the ``privacy barrier,'' as in~\cite{nist}, and used icons---most selected from the Noun Project\footnote{\nounurl}---to represent the different kinds of disclosures. Instead of using an illustration of a database, as in~\cite{nist,Xiong2022UsingIT}, we use an icon of a filing cabinet to represent the collected data. 
Representative diagrams can be found in Appendix~\ref{app:designs}.

\looseness-1 Following guidance from prior work, we also developed privacy labels to clearly demonstrate which kinds of information disclosures DP can protect against. Each row corresponds to a specific information disclosure and clarifies whether protection is offered against said disclosure. We tested three different versions of the tables (six distinct tables in total, across the two models). One version of the table listed only the disclosures against which DP can protect. Thus for the local model, this table had five rows, whereas for the central model, this version had only one row. This table uses a red circle-backslash symbol to indicate that a particular disclosure is not permitted. The other two versions always included information about all five disclosures, but used different iconography to depict protection or lack thereof. 
Both of these versions incorporated lock icons to indicate when DP protected against a particular kind of disclosure. In one of these tables, we use a green lock icon---as recommended in prior work on connection security icons~\cite{felt2016rethinking}---to indicate safety, whereas a red unlocked icon indicates disclosures against which DP does not protect. The other table is in black-and-white and uses the presence or absent of a lock icon to indicate (lack of) protection. Chrome previously used lock icons to indicate connection security, but has recently backed away from this choice due to concerns about overtrust; some Chrome users incorrectly assumed that a lock icon was a reflection on the safety of the website itself rather than the connection~\cite{lockicons,ma2019impact}. Varying the use of icons allowed us to evaluate their appropriateness in a DP context. Appendix~\ref{app:designs} includes representative versions of our original privacy labels. 

\subsection{Protocol}
\label{sec:intproc}
We used a 3 x 2 study design: each participant evaluated either the metaphors, diagrams, or privacy labels for either the local or central model. Our goal was to solicit feedback to help us iterate on our designs of each type. All interviews began by describing the same hypothetical scenario: 

\begin{displayquote}
\textit{A non-profit organization is asking patients around the country to share their medical records, which will be used to help medical research on improving treatment options and patient care. The non-profit would like to explain to people how they will protect patients’ privacy.}
\end{displayquote}

Next, participants are informed that the non-profit plans to ``use an extra layer of privacy protection in order to protect patients’ medical information.'' Then, they are shown the first explanation of this privacy protection. After reading the explanation, the participants are asked to explain how patient data will be protected in their own words, as in~\cite{golla2018site}. Next, they are asked how they feel about the explanation, how well they feel that they understand the privacy protection after reading the explanation, what concerns they would have about sharing their data, and what else they would like to know about how patient data will be protected, adapting questions from~\cite{redmiles2017you}. If the design under discussion includes the use of color, they are also asked about these color choices. Finally, they are asked how the explanation could be improved. 

Next, participants are shown an alternate version of the explanation of the same type, still describing the same model (i.e., local or central). They are asked if the new explanation has changed their understanding. Then, they are asked the same questions they were asked about the original explanation. Some participants were then shown a third version---since we had three versions of the privacy labels and added a third version of the diagrams---and the above questions were repeated. We vary the order of explanations shown between participants. After viewing all explanation versions, participants are asked which one would be most useful for patients deciding whether to share their data. Finally, participants are asked how they would explain to patients how their data would be protected. 

Participants who viewed the privacy label explanations or metaphor explanations were then asked to draw a diagram that conveyed their understanding of how patient data would be protected. Participants who struggled to draw on their screens could choose to tell the interviewer what to draw. The purpose of these drawings is two-fold. The drawings serve both as a way to clarify participants' mental models and as a source of inspiration for iterating on our own designs. The participants who viewed the diagram explanations were not asked to do any drawing, since they would be heavily biased towards the diagrams they had already been shown. Finally, in concluding the interview, participants were asked to self-report gender, race, and ethnicity. Additional demographic information was provided through the recruitment platform.

\subsection{Participant Recruitment}
The first author interviewed 24 U.S. residents recruited through Prolific. We wanted our explanations to be broadly accessible, so we used filters to ensure that at least half of participants had no college degree. A breakdown of participant demographics can be found in Appendix~\ref{app:demographics}. Participants whose interviews included drawing a diagram were paid \$15, whereas participants who evaluated the diagrams were paid \$12 since these interviews were shorter. Interviews lasted about 10-30 minutes and were conducted over Zoom.

\subsection{Analysis}
The interviewer first transcribed and summarized all the interviews. Next, an interview from each condition was selected at random, forming a set of six interviews. The first two authors reviewed these six transcripts to develop a set of codes, organized into four distinct themes (Appendix~\ref{app:codes}). They then shared this codebook with the research team and modified it based on the group's feedback. Finally, the same two authors coded all 24 interviews together\footnote{We do not calculate or report inter-rater reliability (IRR) for two reasons. One, while calculating IRR can be useful to establish agreement before researchers divide a corpus to code different subsets individually, in our case both researchers coded all of the data together. Two, we are not seeking to make quantitative claims about our codes~\cite{McDonald2019}.} with the updated codebook.

\subsection{Findings}
\subsubsection{Effectiveness of Initial Designs}
While we found some strategies more effective than others, across all conditions, participants had additional questions that our explanations did not answer. 

\paragraph{Metaphors.}
Responses to the metaphors were mixed. Some participants appreciated the concision of the metaphors, while others wanted more details. For example, one participant criticized an explanation's brevity, saying it is \textit{``a little bit simple and [...] doesn't go into too many details.'' (P8)} In contrast, a different participant complimented this very quality by describing an explanation as \textit{``reader-friendly, very concise''} (P5). This tension between accuracy and thoroughness of explanations on the one hand, and simplicity on the other has also been reported in other domains, such as explainable machine learning~\cite{abdul2020cogam} and privacy policies~\cite{gluck-16-short-notices}. 

Explanations that make use of metaphor can help people develop useful mental models, and, conversely, people's use of metaphor can reveal their own understanding.  Participants across all conditions provided a range of metaphors conveying their understanding of DP, some of which could be adapted as explanations of DP. For example, a participant in the metaphor condition explained that after their data passed through the privacy barrier, they would be like a ghost, no longer identifiable. Another participant in the metaphor condition explained the obfuscation applied in the local model as follows:
\begin{displayquote}
    \textit{
     I have long hair, but you don't know what color it is. You don't know that I have contacts and not glasses, so you wouldn't be able to pick me out of a lineup, is what I would imagine it as. 
    } (P4)
\end{displayquote}
These metaphors of ghosts and lineups both hint at the idea of DP as a form of anonymization. 
This same participant provided another particularly creative metaphor: 
\begin{displayquote}
    \textit{
    It's kind of like an egg. You know, you crack it open and you don't know if it's going to be rotten inside or not. But I don't know what chicken it came from, so I can't blame the chicken. 
    } (P4)
\end{displayquote}
The phrase ``can't blame the chicken'' seems to convey the protection offered by DP as a form of plausible deniability. 

\begin{mdframed}[skipabove=\topskip]
\noindent
\textbf{Design Changes:}
We replaced our original metaphors with a new metaphor inspired by those generated by participants. Synthesizing metaphors related to hiding or changing one's appearance---like not being recognizable in a lineup or becoming a ghost---we developed a new metaphor: this metaphor compares protecting data with DP to wearing a ``disguise.'' 
\end{mdframed}

\paragraph{Diagrams.}
\looseness-1 Of all the explanation methods, the diagrams were the least successful. Of the eight participants assigned to this condition, five explicitly expressed that the diagram was confusing. Although the other three participants did not explicitly use the term ``confusing,'' they also struggled to understand various aspects of the diagrams. For example, when asked to explain the privacy protection in their own words, one participant started to try to explain, then cut themselves off and responded: \textit{``Well, I don't really know''} (P23).

A number of participants expressed confusion or disagreement with the underlying threat model, particularly for the central model. The central model only prevents disclosure from published reports. Although responsible data collectors will employ other technologies such as encryption to protect against hackers or criminals, DP in itself does not protect against this kind of disclosure in the case of the central model. For some participants, this was counterintuitive. For example, after viewing a diagram explaining the central model, one participant expressed their confusion as follows: 
\begin{displayquote}
\textit{
     I don't really get it. [$\cdot\cdot\cdot$] There's supposed to be a barrier between my medical information and the people who read the published reports. It seems. And then people who want your data seems like that's open and free, and it seems backwards to me.
} (P24)
\end{displayquote}
Two other participants viewing diagrams for the central model incorrectly stated that the privacy barrier was protecting data from hackers, even though the diagrams showed hackers on the left side of the privacy barrier (i.e., the same side as the data collection). One of these participants realized their mistake later in the interview. First, they explained:
\begin{displayquote}
    \textit{The privacy barrier [$\cdot\cdot\cdot$] allows the people who utilize the information, say the law enforcement and medical professionals, [$\cdot\cdot\cdot$] to share that information amongst themselves on a secure in a secure network without allowing the people who want to get that information to abuse that information, the hackers}. (P22)
\end{displayquote} 
However, a bit later, they realized their mistake:
\begin{displayquote}
    \textit{
     I'm looking at it again. It says well the people want that data, it's just letting them take it, it looks like. So I guess that would kind of be a concern there [$\cdot\cdot\cdot$] we're letting the scientists and the policymakers, the scientists, the people who need to see maybe medical data not allowing them to see the data. But it has a backdoor that allows the people who want to steal that information. So it really has a flaw. 
    } (P22)
\end{displayquote}
Despite the fact that the diagrams showed the hacker to the left of the privacy barrier, two of the four participants in this condition nevertheless explicitly stated that the privacy barrier would protect their data from hackers. Many people may expect PETs to protect against hackers and criminals, making the protection offered by central DP alone somewhat unintuitive~\cite{smart2022}. We dropped the diagram explanations due to the pervasive confusion expressed by participants.

Xiong et al.~\cite{Xiong2022UsingIT} previously investigated the use of diagrams for explaining location privacy and found less than ideal levels of comprehension, particularly for the local model, though they speculate that data quality issues with Amazon Mechanical Turk may be to blame. Alternatively, it is possible that data flow diagrams inherently overemphasize \textit{processes} at the expense of clearly enumerating \textit{implications}. 

\begin{mdframed}[skipabove=\topskip]
\noindent
\textbf{Design Changes:}
The diagram explanations were dropped, due to persistent confusion.
\end{mdframed}

\paragraph{Privacy Labels.}
Responses to the privacy labels were largely positive, though not universally so. Participants praised the privacy labels for their simplicity and clarity. In addition, several participants appreciated the use of color. For example, one participant explained that: ``\textit{Having the colored icons does make it a bit faster for a person to get the message}'' (P16). However, participants did not always agree about the meaning of the colors green and red. On the one hand, green is often associated with safety while red is associated with danger. Given these associations, one might use green to indicate protection and red to indicate vulnerability. On the other hand, green is also used to mean ``go'' whereas red means ``stop.'' Given these associations, one might use red to indicate protection, since the flow of data is ``stopped.'' Some participants felt that our use of green and red should be switched, while others felt that our use was appropriate.

\begin{mdframed}[skipabove=\topskip]
\noindent
\textbf{Design Changes:}
To ameliorate the confusion with red and green, we eliminated red and chose to highlight protection in green. The rest of the content was black.  
\end{mdframed}

\paragraph{Importance of Process.}
All of our explanations were designed to communicate the \textit{implications} of DP rather than the details of \textit{how} DP works. Prior work has shown that explaining the process of adding noise to data is not enough to help people understand the consequences data sharing~\cite{xiong2020towards}. Nevertheless, omitting any discussion of process seems to leave people unsatisfied and confused. Most participants had questions about how the data protection worked. Providing a detailed explanation of the mathematical and technical details of DP is likely to overwhelm most people, but people nevertheless do want some information about how DP works---finding the right balance may be challenging. This finding aligns with prior work on explaining encryption. While explanations of encryption focused on \textit{outcome} lead to greater perceived security than explanations focused on \textit{process}, hybrid explanations that incorporate information on both process and outcome lead to the greatest perceived security~\cite{distler2020making}. Prior work on metaphors for DP also found that some participants were interested in understanding how DP works~\cite{karegar2022exploring}.

\begin{mdframed}[skipabove=\topskip]
\noindent
\textbf{Design Changes:}
Another text was added to provide context about how DP works; we adapted a state-of-the-art text explanation by Xiong et al.~\cite{xiong2020towards}, while aiming for improved readability by eliminating terms like ``database'' and ``aggregated.'' We anticipated that this additional information on \textit{process} could complement our other explanations that focus on \textit{implications}.
\end{mdframed}

\subsubsection{Mental Models}
\label{sec:interview:mentalmodels}
Our interviews reveal a number of different mental models that participants constructed to understand DP, based on the explanations they were shown. In many cases, participants' mental models were informed by their prior knowledge of and experience with other technologies. 

\paragraph{Comparison to other PETs.}
Some participants---especially in the diagram 
and privacy label 
conditions---reasoned about DP through comparisons to other PETs. For example, one participant understood the privacy barrier as \textit{``some kind of firewall that keeps [their] privacy safe''} (P21). Encryption in particular was mentioned frequently, perhaps because it is a particularly familiar and ubiquitous PET or perhaps because participants associated our lock icons with encryption~\cite{felt2016rethinking,herzberg2016can}. One participant, who assumed that encryption was the technology being described, wanted to know \textit{``what type of encryption''} (P23) was used. Prior work has also found associations between DP and encryption, and found that associations with encryption correspond to higher trust~\cite{karegar2022exploring}. DP is distinct from encryption, so while it may be possible to leverage people's knowledge about encryption to construct better explanations of DP, it is also likely that associations with encryption may lead to misconceptions. 

One particular source of confusion is that with encryption, the protection offered should be binary---information is either encrypted (i.e., protected) or not. This corresponds nicely with the physical metaphor of a lock that has exactly two states: locked and unlocked. In the case of DP, however, the goal is to allow some information ``leakage'' while still offering some protection---the amount of information leakage depends on the the privacy budget parameter. Although many participants liked the lock icons, other participants pointed out this issue. For example, one participant in the diagram condition said:
\begin{displayquote}
    \textit{If you're releasing some form of my information to these published reports, it's not completely locked. 
} (P20)
\end{displayquote}
Thus, the use of lock icons and their association with encryption may in some cases prove problematic. 

\begin{mdframed}[skipabove=\topskip]
\noindent
\textbf{Design Changes:}
We designed an additional version of the privacy labels that uses arrows to indicate whether data flows are permitted or blocked instead of locks. This version uses red to denote flows that are blocked. 
\end{mdframed}

\paragraph{DP as anonymization.}
Several participants understood DP as an anonymization technique---especially those who read the metaphor explanations. 
These participants often had an overly-simplistic view of DP. For example, one participant explained that in their understanding, the data \textit{``would be protected by virtue of being anonymized and not including the patient's name, social security number, or date of birth''} (P13). Of course, DP provides better guarantees than such a naive anonymization strategy; nevertheless, this mental model may provide a useful approximation of practical DP guarantees.

\paragraph{DP as fake data.} A few participants understood DP as the injection of fake data. One participant explained it as follows:
\begin{displayquote}
    \textit{
    You're collecting my name, but it's a fake one, so it's like a shield up in front of me.
    } (P4)
\end{displayquote}
Once again, while this model oversimplifies DP, 
it shares key elements with the truth and thus is likely useful overall. However, it is important for people to understand that the ``fake'' data nonetheless reveals useful information about the overall distribution; therefore, DP does not necessarily offer protection against inferential privacy risks~\cite{kifer2011no,kifer2014pufferfish}.

\subsubsection{Validating Design Changes.}
\looseness-1 We recruited 10 additional participants through Prolific to pilot our updated explanations. These participants were shown explanations of various types---including the two privacy labels and various texts that evolved somewhat over the course of the interviews---and asked to build their own explanation by editing or combining existing explanations or creating their own from scratch (Figure~\ref{fig:miro}). Participants expressed more satisfaction and few substantive edits as compared to our initial evaluations, however they suggested a wide range of ways to combine the texts and privacy labels. No singular combination was preferred by several participants. Therefore, in our quantitative evaluation, we test not only the texts and privacy labels alone but also these explanations in combination with each other as further detailed in Section~\ref{sec:eval}. 

Further, prior to launching the quantitative evaluation of our designs, we compared our two privacy labels in a survey using the evaluative criteria outlined in Section~\ref{sec:eval}. We found no significant differences between the two versions on any of the evaluation criteria. We chose to continue with the version with arrows instead of the version with locks for a few reasons. One participant expressed their preference for the version with arrows over the one with locks as follows:
\begin{displayquote}
    \textit{
     I felt better seeing the same people being blocked rather than the lock because you see those everywhere nowadays.
    } (P28)
\end{displayquote} In other words, the lock symbol has become so ubiquitous that this participant found it meaningless. We also felt that the arrows more clearly showed that certain information disclosures were protected against while others were not, whereas lock icons might suggest that certain people are given a ``key.'' This is a fundamentally different kind of protection since keys can be leaked or shared. 
Finally, although the difference was not significant, comprehension scores were slightly better for the version with arrows. Thus, we dropped the version with locks. The evolution of our designs is visualized in Figure~\ref{fig:flow}.
\begin{mdframed}[skipabove=\topskip]
\noindent
\textbf{Design Changes:}
We dropped the label with locks in favor of the version that emphasized information flows.
\end{mdframed}

%% file: 4_survey.tex
We conducted an online survey in March 2023 to evaluate the impact of our explanations on understanding and to assess their efficacy in setting appropriate privacy expectations.

\begin{wrapfigure}[22]{r}{0.5\textwidth}
\center 
\includegraphics[width=0.45\textwidth]{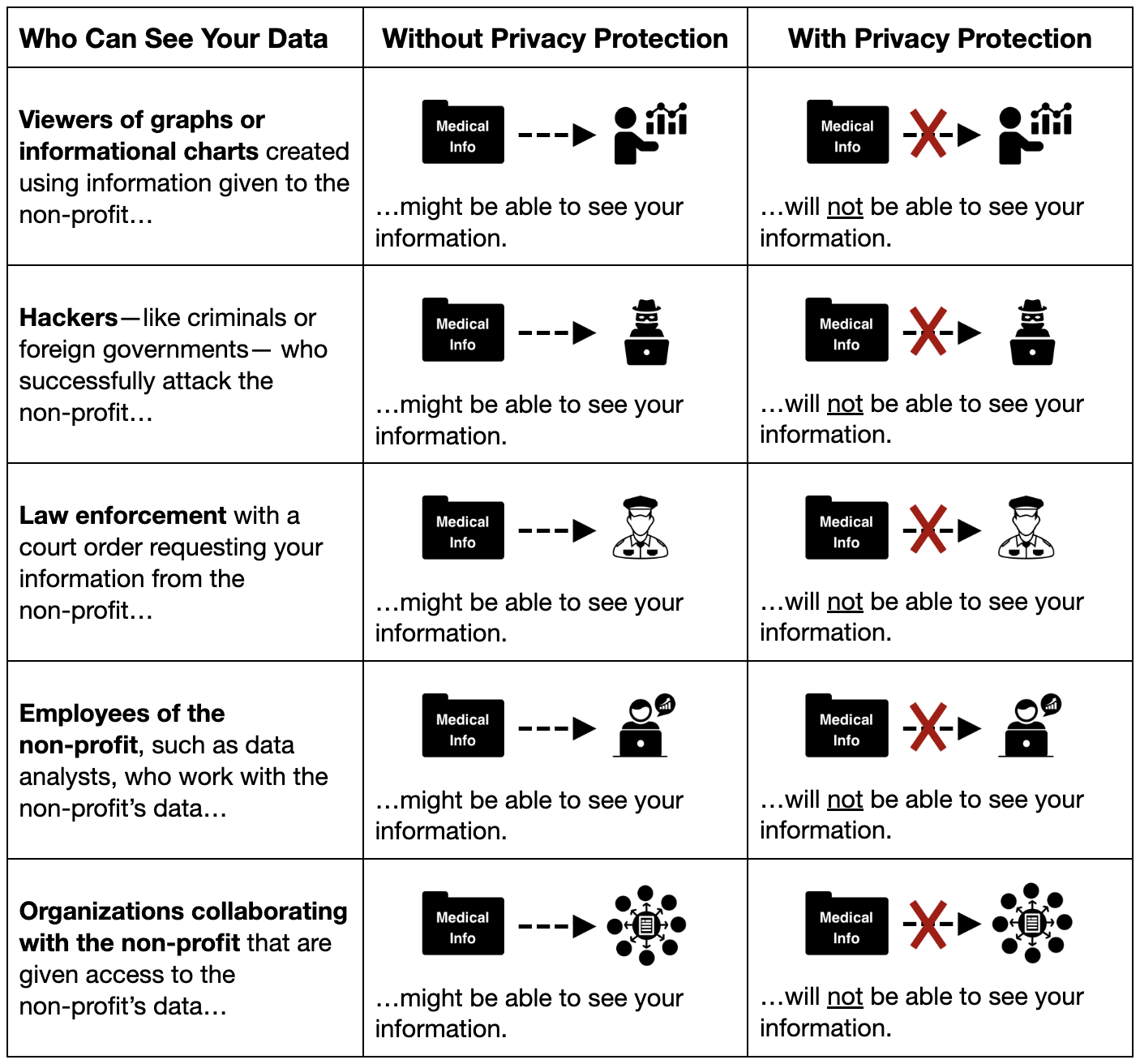}
\caption{\label{fig:label} The final version of our privacy label for the local model. The central version is included in Appendix~\ref{app:designs}.}
\end{wrapfigure}

\paragraph{Protocol.}
Respondents are first asked to read the scenario description (the same medical scenario discussed in Section~\ref{sec:intproc}) and the description of how data will be protected. Next, respondents answer a simple, multiple-choice comprehension question to ensure that they have read the scenario description. They are given the option to re-read the description. If they do not answer correctly, they are given a second attempt, in accordance with Prolific's policies. If after the second attempt, they again answer incorrectly, they are prevented from advancing further in the survey. \looseness=-1

Respondents who pass the comprehension check are then asked whether they trust the non-profit to protect patient privacy~\cite{xiong2020towards}. Next they are asked two questions related to self-efficacy. Finally, they are asked whether they would be willing to share their information with the non-profit. An open-text box asks them to explain their decision. 

Respondents then answer five true/false questions on privacy expectations, followed by the Likert-scale questions about understanding and thoroughness. They are also invited to share feedback on the explanations in a free-response text box. When answering the above questions, respondents have the option to reread the descriptions of the scenario and privacy protection at any time. Next, respondents are asked about their familiarity with various PETs, including DP and a non-existent technology (``deliquescent security''). If they indicate familiarity with some of the listed technologies, they are asked which of the technologies (if any) was described in the survey. In a free-response text box, they are asked to explain their reasoning. Finally, respondents answer questions about themselves. In addition to standard demographic questions (i.e., age, income, race, ethnicity, gender, education, job field), the survey also includes measures of internet skill~\cite{hargittai2012succinct}. 
The full survey instrument is included in Appendix~\ref{app:survey}. \looseness=-1

\input{tables/explanations}

\paragraph{Experimental Conditions.}
 We use an 8 x 2 experimental design (all conditions listed in Appendix~\ref{app:designs}), varying both the explanation type and the deployment model---local or central. We evaluate the privacy labels (Figure~\ref{fig:label}), process text, and metaphor text individually as well as in combination: metaphor + process, metaphor + process + label, metaphor + label, and process + label. We compare these seven conditions against the implication-focused explanations from~\cite{xiong2020towards}. Although~\cite{xiong2020towards} evaluated a number of different explanations, we chose to compare against the explanation that led to the highest comprehension of privacy protections. Table~\ref{tab:conditions} provides all three text explanations.

\paragraph{Dependent Measures.}
\label{sec:dvs}
Our goal is to set privacy expectations appropriately. 
Thus, we use a series of true/false questions from prior work about whether certain types of disclosure are possible to measure \emph{objective comprehension} (Table~\ref{tab:disclosures}). 
We also ask respondents about their \emph{subjective understanding} of the explanations and how \emph{thorough} they perceive the explanations to be~\cite{kelley2009nutrition}.
Additionally, we ask whether respondents \emph{trust} the non-profit organization to protect patient privacy~\cite{xiong2020towards}, and we ask two questions related to \emph{self-efficacy} in decision making~\cite{nanayakkara2023chances}. All five questions use 5-pt Likert scales. Finally, although we do ask about \emph{willingness to share} data with the non-profit, we caution against using this as a measure of explanation quality. The explanation that convinces the most people to share their data is not necessarily the best explanation. For example, we hope that a patient who is particularly concerned about disclosure to law enforcement would choose \textit{not} to share data when it is protected using the central model.

\paragraph{Participant Recruitment.}
698 total respondents were recruited through Prolific, using the ``balanced sample'' feature---in accordance with best practices---to recruit an approximately representative sample in terms of gender~\cite{tang2022well}. We conducted a power analysis to estimate an appropriate sample size; due to the large number of experimental conditions, we lack the statistical power to detect very small effects, but such small effects are unlikely to be meaningful in real-world contexts~\cite{CCS:RZKKDM18}. Respondents were paid \$2 for completing the survey, and the median completion time was just under six minutes.
A detailed breakdown of respondent demographics can be found in Appendix~\ref{app:demographics}. 

\paragraph{Analysis.}
We analyze\footnote{Analysis code: {\tiny \myurl}} the effect of our explanations on our dependent measures. We construct a set of regression models studying the effect of our independent variables---explanation and model---on our dependent variables: objective comprehension, subjective understanding, perceived thoroughness, trust, self-efficacy, and data-sharing decision. We use logistic regression models to study data-sharing decisions, linear regression models to study comprehension, and ordinal regression models to study the remaining dependent variables. In all models, we control for internet skill~\cite{hargittai2019internet}. 

We also perform a qualitative analysis of the responses to two of the free-response questions. The first author reviewed all the reasons respondents gave for their data-sharing decisions and developed a set of codes. The first and second authors then reviewed the codebook together and coded 30 responses, resolving disagreements through discussion and refining the codebook as necessary. Then they separately coded 25 responses and evaluated inter-rater agreement by calculating Cohen's Kappa---the average across all codes appearing in this sample was 0.75, indicating substantial agreement. Remaining responses were divided between both authors for coding. After finding the privacy labels to be most effective, the first author additionally reviewed all feedback provided for the privacy labels and developed a second set of codes---despite some overlap in the themes discussed in the feedback responses and the data-sharing decision responses, the content was sufficiently distinct to merit separate codebooks.  
Again, the first two authors reviewed the codebook and coded 10 responses together, resolving disagreements through discussion. Then they separately coded 25 responses and calculated Cohen's Kappa, with an average of 0.98 across all codes appearing in the sample, indicating near perfect agreement. The remaining 393 responses from participants in any of the privacy label conditions were divided between both authors for coding. For both sets, multiple codes could be applied to a single response. Both sets of codes are available in Appendix~\ref{app:codes}. 

\input{tables/understanding}

\subsection{Results}
\subsubsection{Effectiveness of Designs}
We find some explanations are more effective than others in terms of objective comprehension and trust.

\paragraph{Comprehension.}
\looseness-1 Across all explanations, we find a significant difference in objective comprehension (Table~\ref{tab:comprehension}) between the local and central model---the local model is associated with fewer correct answers (\linReg{-1.17}{<0.01}). 
This finding is consistent with prior work which suggests that privacy expectations are more closely aligned with the central model than with the local model~\cite{CCS:CumKapRed21, Xiong2022UsingIT}. It may be difficult to realign a reader's understanding if they come in with strong expectations that do not match the actual protection offered by DP. 
Compared to the Xiong et al. explanation, all of the explanations that include a privacy label are associated with more correct answers (\linReg{0.95 \text{--} 1.26}{< 0.01}). The text-only explanations, on the other hand, showed no significant improvement over the Xiong et al. explanation. This improvement is expected since the privacy labels are designed explicitly to highlight the information flows that respondents are asked about in the comprehension questions. Interestingly, there is a misalignment between objective comprehension and subjective understanding. The process+metaphor explanation is the only one that significantly improves subjective understanding compared to the Xiong et al. baseline (\ordReg{1.79}{<0.05}), even though it does not improve objective comprehension. 
Indeed, objective comprehension and subjective understanding are not strongly correlated ($\tau=0.085; p<0.01$), though there is stronger correlation for the local model ($\tau=0.159; p<0.001$) than for the central model ($\tau=-0.0389; p>0.1$). Prior work has found similar misalignment between objective comprehension and subjective understanding~\cite{smart2022,CCS:FVSTM22}. Unsurprisingly, internet skill is also associated with higher objective comprehension (\linReg{0.20}{<0.01}) and subjective understanding (\ordReg{1.36}{<0.001}). 

\input{tables/dvs}
\input{tables/share}

\paragraph{Other Evaluation Criteria.}
Table~\ref{tab:dvs} summarizes how the explanations compare on our other evaluation criteria. Although comprehension is better for the central model, trust is higher for the local model  (\ordReg{1.97}{<0.05}). This is promising, since the local model does offer stronger privacy. The label + process explanation is also associated with greater trust. 
This aligns with the qualitative feedback from our interviews. While information about process is not enough to help readers understand implications, it seems that explanations that focus only on implications leave readers feeling skeptical. This result is consistent with prior work on explaining encryption that finds benefits of combining information on process and outcome~\cite{distler2020making}. There were no significant effects of model or explanation on perceived thoroughness or self-efficacy, although higher internet skill is associated with higher self-efficacy (\ordReg{1.21 \text{--} 1.28}{<0.05}).

\paragraph{Feedback.}
As in our interview study, one of the most common themes in our respondents' feedback was a desire for more information about how the privacy protection works (n=76). For example, 
one respondent wrote:
\begin{displayquote}
    \looseness-1 \textit{It doesn’t explain at all how this supposed ``privacy protection'' works, so how do I know if it’s credible? I have a lot of cybersecurity training: I want technical details! }
\end{displayquote}
Even respondents who read the process text sometimes requested more information about data protection processes. Respondents also requested other kinds of additional information (n=28), for example, about the organization and how it would use their data.
A tension was again evident between respondents who requested additional information and those who praised our concision or requested further simplification. One respondent suggested \textit{``more detailed explanations of the privacy protections that are available to view if needed.''}
This adaptive approach was also suggested in interviews.

\subsubsection{Prior Familiarity with PETs}
In interviews, we found that some participants understood DP through comparisons with other PETs. 
Of the PETs we mention in our survey, end-to-end encryption was by far the most familiar, whereas only a minority had heard of DP (Appendix~\ref{app:demographics}). Of respondents who answered the question asking which technology was described in the survey, most correctly selected DP, though several respondents explained in their free-text responses that they were simply guessing.

\subsubsection{Data-Sharing Decision}
Respondents are more willing to share data (Table~\ref{tab:share}) under the local model (\ordReg{1.46}{<0.05}). This replicates findings from prior work and is likely due to the fact that the local model offers stronger privacy guarantees~\cite{xiong2020towards}. None of the explanations had a significant effect on data-sharing decisions. 

When people are deciding whether to share information, they consider many other factors in addition to privacy protections~\cite{smart2022,naeini2017privacy,frik2023model}. In fact, many respondents simply were not worried about privacy (n=75). For example, one respondent felt that they had nothing in their medical history that they would ``\textit{need to hide or be particularly private about}.'' Other respondents were interested in sharing their information, because they value helping others, mentioning benefits of data sharing (n=151). In the words of one respondent: ``\textit{I do not have a problem with sharing my records if it will help someone}.'' On the other hand, respondents who were less willing to share their data often indicated that they felt it would be too risky or that their medical information was simply too private (n=242). For example, one respondent explained they were ``\textit{not comfortable sharing [their] medical records with anyone but [their] doctor}.'' Other respondents wanted more information before they would be willing to share their data (n=155). However, the information they requested was not always related to DP. For example, some respondents wanted to know more about the non-profit organization. Finally, some participants distrusted either the non-profit or the privacy protection (n=88). In the words of one respondent:
\begin{displayquote}
    \textit{Companies say that your information is secure all the time, but all the time there are security breaches. I do not trust my private information to be secure with anyone.}
\end{displayquote}
Other respondents also mentioned the frequency of data breaches as a cause for concern (n=35).

%% file: tables/explanations.tex
\begin{table*}[t]
  \caption{Explanation texts for the local and central models.}
  \label{tab:conditions}
  \begin{tabular}{l m{6.9cm} m{6.9cm}}
    \toprule
    Type&Local&Central\\
    \midrule
    Process
    & 
    {\small  
    To protect your information, your data will be randomly modified before it is sent to the organization. Only the modified version will be stored, so that your exact data is never collected by the organization.
    } 
    &
    {\small 
    To protect your information, the organization will store your data but only publish reports, graphs, or charts that have been randomly modified. These modifications hide information that is unique to you as an individual.
    }
    \\
    \midrule
   Metaphor
   & 
   {\small 
   The technology works something like this: Your data will be disguised before it is stored by the organization. Therefore, anyone who accesses the data collection will only see this disguised version of your data.
   }  
   &
   {\small 
   The technology works something like this: The collected data will be disguised when any graphs, charts, or reports are published. However, anyone who accesses the organization’s data collection will see the undisguised data.
   }
   \\
   \midrule
   Xiong et al.
   & 
   {\small 
   To respect your personal information privacy and ensure best user experience, the data shared with the non-profit organization will be processed via an additional privacy technique. That is, your data will be randomly modified before it is sent to the organization. Since the organization stores only the modified version of your personal information, your privacy is protected even if the organization's database is compromised.
   } 
   &
   {\small 
   To respect your personal information privacy and ensure best user experience, the data shared with the non-profit organization will be processed via an additional privacy technique. That is, the organization will store your data but only publish the aggregated statistics with modification so that your personal information cannot be learned. However, your personal information may be leaked if the organization’s database is compromised.
   }
   \\
  \bottomrule
\end{tabular}
\end{table*}

%% file: tables/understanding.tex
\begin{table*}[h]
\centering  
\begin{tabular}{llr|lr} 
 \toprule
 \multicolumn{1}{c}{\textbf{Variable}} 
 & \multicolumn{2}{c|}{\textbf{\emph{Objective Comprehension}}} 
 & \multicolumn{2}{c}{\textbf{\emph{Subjective Understanding}}} 
  \\
 

\midrule
& $\beta$ & CI & OR & CI  \\ 

 \midrule

 {\small Model: Local} & {\small $\mathbf{-1.17}^{***}$} & {\scriptsize $[-1.42, -0.92]$} & {\small $0.79$} & {\scriptsize $[0.61, 1.03]$} \\ 
 
 {\small Expl: Metaphor} & {\small $0.09$} & {\scriptsize $[-0.45, 0.62]$} & {\small $1.63$} & {\scriptsize $[0.91, 2.90]$} \\ 
 
 {\small Expl: Process} & {\small $-0.33$} & {\scriptsize $[-0.86, 0.19]$} & {\small $1.39$} & {\scriptsize $[0.79, 2.45]$} \\ 

 {\small Expl: Process+Metaphor} & {\small $0.47$} & {\scriptsize $[-0.06, 1]$} & {\small $\mathbf{1.79}^{*}$} & {\scriptsize $[1.01, 3.18]$} \\ 

 {\small Expl: Arrow Label} & {\small $\mathbf{1.15}^{***}$} & {\scriptsize $[0.62, 1.68]$} & {\small $1.35$} & {\scriptsize $[0.75, 2.42]$} \\ 


 {\small Expl: Label+Metaphor} & {\small $\mathbf{1.26}^{***}$} & {\scriptsize $[0.73, 1.8]$} & {\small $1.51$} & {\scriptsize $[0.86, 2.67]$}\\ 
 
 {\small Expl: Label+Process} & {\small $\mathbf{0.97}^{***}$} & {\scriptsize $[0.45, 1.5]$} & {\small $1.39$} & {\scriptsize $[0.79, 2.44]$} \\ 

 {\small Expl: Label+Process+Metaphor} & {\small $\mathbf{0.95}^{***}$} & {\scriptsize $[0.43, 1.48]$} & {\small $1.48$} & {\scriptsize $[0.84, 2.6]$} \\ 

 {\small Internet Skill} & {\small $\mathbf{0.20}^{**}$} & {\scriptsize $[0.05, 0.35]$} & {\small $\mathbf{1.36}^{***}$} & {\scriptsize $[1.16, 1.6]$} \\ 

\midrule
\end{tabular}
\caption{ \small \label{tab:comprehension} \emph{Left:} results from linear regression models for objective comprehension.
We report regression coefficients ($\beta$) and 95\% CIs for these coefficients. $\beta>0$ indicates an increase while $\beta<0$ indicates a decrease.
\emph{Right:} results from ordinal regression models for subjective understanding.
We report odds ratios (OR) and corresponding 95\% CIs. An OR $>1$ indicates an increase in odds, while an OR $<1$ indicates a decrease. For both columns, we use the Xiong et al. explanation as the reference level explanation.* p$<0.05$; ** p$<0.01$; *** p$<0.001$.}
\end{table*}

%% file: tables/dvs.tex
\begin{table*}[h]
\centering   
\begin{tabular}{llr|lr|lr|lr} 
 \toprule
 {\textbf{Variable}} & \multicolumn{2}{c|}{\textbf{\emph{Trust}}} & \multicolumn{2}{c|}{\textbf{\emph{Thoroughness}}}
  & \multicolumn{2}{c|}{\textbf{\emph{SE (Info)}}} & \multicolumn{2}{c}{\textbf{\emph{SE (Confidence)}}} \\
 

\midrule
& OR & CI & OR & CI & OR & CI & OR & CI \\ 

 \midrule

 {\small Model: Local} & {\small $\mathbf{1.70}^{***}$} & {\scriptsize $[1.3, 2.23]$} & {\small $1.08$} & {\scriptsize $[0.83, 1.41]$} & {\small $0.87$} & {\scriptsize $[0.67, 1.13]$} & {\small $0.90$} & {\scriptsize $[0.69, 1.17]$} \\ 
 
 {\small Expl: Metaphor} & {\small $1.46$} & {\scriptsize $[0.82, 2.6]$} & {\small $1.28$} & {\scriptsize $[0.72, 2.27]$} & {\small $1.21$} & {\scriptsize $[0.67, 2.19]$} & {\small $1.65$} & {\scriptsize $[0.91, 2.99]$} \\ 
 
 {\small Expl: Process} & {\small $0.99$} & {\scriptsize $[0.56, 1.73]$} & {\small $0.81$} & {\scriptsize $[0.46, 1.43]$} & {\small $0.70$} & {\scriptsize $[0.39, 1.23]$} & {\small $0.95$} & {\scriptsize $[0.53, 1.68]$} \\ 
 
{\small  Expl: Process+Metaphor} & {\small $1.45$} & {\scriptsize $[0.81, 2.58]$} & {\small $1.55$} & {\scriptsize $[0.88, 2.74]$} & {\small $0.93$} & {\scriptsize $[0.52, 1.64]$} & {\small $1.19$} & {\scriptsize $[0.67, 2.11]$} \\ 
 
 {\small Expl: Arrow Label} & {\small $0.85$} & {\scriptsize $[0.48, 1.52]$} & {\small $0.68$} & {\scriptsize $[0.38, 1.21]$} & {\small $1.15$} & {\scriptsize $[0.64, 2.05]$} & {\small $1.44$} & {\scriptsize $[0.8, 2.61]$} \\ 
 
 
 {\small Expl: Label+Metaphor} & {\small $1.18$} & {\scriptsize $[0.66, 2.12]$} & {\small $1.73$} & {\scriptsize $[0.97, 3.09]$} & {\small $0.96$} & {\scriptsize $[0.53, 1.72]$} & {\small $1.16$} & {\scriptsize $[0.65, 2.06]$} \\ 
  
 {\small Expl: Label+Process} & {\small $\mathbf{1.97}^{*}$} & {\scriptsize $[1.12, 3.47]$} & {\small $1.22$} & {\scriptsize $[0.7, 2.15]$} & {\small $1.08$} & {\scriptsize $[0.62, 1.87]$} & {\small $1.06$} & {\scriptsize $[0.61, 1.87]$} \\ 
 
 {\small Expl: Label+Process+Metaphor} & {\small $0.94$} & {\scriptsize $[0.54, 1.64]$} & {\small $1.38$} & {\scriptsize $[0.79, 2.41]$} & {\small $0.98$} & {\scriptsize $[0.55, 1.73]$} & {\small $1.07$} & {\scriptsize $[0.6, 1.9]$} \\ 
 
 {\small Internet Skill} & {\small $0.96$} & {\scriptsize $[0.82, 1.13]$} & {\small $1.06$} & {\scriptsize $[0.9, 1.25]$} & {\small $\mathbf{1.21}^{*}$} & {\scriptsize $[1.03, 1.41]$} & {\small $\mathbf{1.28}^{**}$} & {\scriptsize $[1.09, 1.5]$} \\ 
 
\midrule
\end{tabular}
\caption{ \small \label{tab:dvs} Results from regression models for trust, perceived thoroughness, and self-efficacy, with the Xiong et al. explanation as the reference level explanation. Again we report odds ratios with 95\% CIs. 
An OR $>1$ indicates an increase in odds, while an OR $<1$ indicates a decrease.
}
\end{table*}

%% file: tables/share.tex
\begin{wraptable}[18]{r}{0.55\textwidth}
\centering  
\begin{tabular}{llr} 
 \toprule
 {\textbf{Variable}} & \multicolumn{2}{c}{\textbf{\emph{Share}}} \\
 

\midrule
& OR & CI \\ 

 \midrule

 {\small Model: Local} & {\small $\mathbf{1.46}^{*}$}   
 & {\scriptsize $[1.07, 2]$}  \\
 
 {\small Expl: Metaphor} & {\small $1.09$}   
 & {\scriptsize $[0.57, 2.12]$}  \\
 
 {\small Expl: Process} & {\small $0.75$}   
 & {\scriptsize $[0.38, 1.45]$} \\

 {\small Expl: Process+Metaphor} & {\small $0.81$}   
 & {\scriptsize $[0.41, 1.58]$} \\

 {\small Expl: Arrow Label} & {\small $0.52$}   
 & {\scriptsize $[0.26, 1.04]$} \\


 {\small Expl: Label+Metaphor} & {\small $0.80$}   
 & {\scriptsize $[0.4, 1.56]$} \\
 
 {\small Expl: Label+Process} & {\small $1.33$}   
 & {\scriptsize $[0.7, 2.56]$} \\

 {\small Expl: Label+Process+Metaphor} & {\small $0.65$}   
 & {\scriptsize $[0.33, 1.28]$} \\

 {\small Internet Skill} & {\small $0.95$}   
 & {\scriptsize $[0.79, 1.15]$} \\

\midrule
\end{tabular}
\caption{ \small \label{tab:share} Results from regression model for data-sharing decision. We report odds ratios (OR) and corresponding 95\% CIs. An OR $>1$ indicates an increase in odds, while an OR $<1$ indicates a decrease.}
\end{wraptable}

 



 
 




 




%% file: 5_limitations.tex
Our designs are limited in their focus on a single scenario. Although medical applications are often cited as motivation for studies of DP~\cite{jordon2018pate, bhaskar2010discovering}, DP has not been widely deployed in medical contexts~\cite{dankar2013practicing}. Nevertheless, our privacy labels are transportable to other domains. Future work could transfer our designs to other scenarios and test whether our findings still hold. A limitation of our evaluation is that encountering explanations of DP in practice differs significantly from encountering explanations in an online survey. Future work could investigate comprehension when these explanations are encountered in more natural settings. A third limitation is our focus on a U.S. audience. Our privacy labels may be received differently in a different cultural context. Finally, we present the nature of DP's protection as binary, when in fact the level of protection depends on the choice of privacy budget. This simplification may be appropriate for small privacy budgets, but the question of determining an acceptable range for the privacy budget is itself a nontrivial problem.

One concern may be that our privacy labels are ``teaching to the test,'' since we design them specifically to highlight the information disclosures that we ask about to measure comprehension. Thus, it is not surprising that comprehension is higher for our privacy labels than for explanations designed with a different emphasis. However, if the purpose of an explanation is to inform readers about which information flows are restricted---i.e., if we are using the ``right'' test---perhaps teaching to the test is not such a problem. Nevertheless, we incorporate additional evaluation criteria from prior work and find that our privacy labels improve comprehension without sacrificing quality on these other metrics (Tables~\ref{tab:comprehension}--~\ref{tab:dvs}).

%% file: 6_discussion.tex
Our results highlight the value of combining disparate best practices from prior work on explaining other security and privacy (S\&P) concepts to explain complex PETs such as DP~\cite{kelley2009nutrition,distler2020making}. We find that consequences-focused explanations (i.e., privacy label explanations that highlight information flows) to be a promising approach for promoting accurate understandings of potential data leaks in DP systems. However, to ensure that such explanations are trusted we find that it is necessary to pair such consequences-focused information with a limited amount of high-level information about mechanisms: how DP works to offer particular consequences and protections. 
Below we discuss potential pitfalls of privacy labels for DP as well as ways to extend our designs to explain other PETs individually or in combination.

\paragraph{Potential Pitfalls.}
Although the nutrition label approach shows promise for setting appropriate privacy expectations, it is important to avoid pitfalls from prior deployments of nutrition labels for privacy~\cite{cranor2022mobile}. For example, iOS privacy labels can be misleading and inaccurate~\cite{kollnig2022goodbye}, in part because developers struggle to create accurate labels~\cite{li2022understanding}. Similarly, our labels for DP could be misleading if an organization has implemented DP incorrectly~\cite{jin2022,Bichsel2021,CCS:CSVW22,CCS:Mironov12,lyu2017} or has chosen an inappropriately large privacy budget~\cite{dwork2019differential}. Specialized programming platforms, audits, and formal verification approaches are therefore an important complement to our work~\cite{mcsherry2009, reed2010, zhang2017, tramer2022debugging,CCS:DWWZK18,kifer2020guidelines}, namely by ensuring that the communicated privacy guarantees match the implementation. 

\looseness-1 Furthermore, while privacy labels can empower individuals to make decisions that better align with their goals and values, it is also important not to overburden individuals in the same way that traditional privacy policies do~\cite{mcdonald2008cost}. As some of the participants we interviewed highlighted, it can be difficult to strike the right balance between simplicity and comprehensiveness. Such a balance is important not only for data subjects, but also for other audiences who may encounter DP. 
For instance, privacy labels for DP could be used to educate policymakers, advocacy organizations, or software developers to support them in various decision-making processes. For example, Mozilla's ``privacy not included'' guide offers expert reviews to help buyers choose products that provide strong privacy and security, since it can be difficult for individual buyers to evaluate various data protection policies themselves. One could imagine a similar project to provide reviews for different data collection initiatives. An advocacy organization might use privacy labels for PETs like DP to identify and recommend certain initiatives that provide good S\&P guarantees.

Finally, it is crucial that privacy labels for DP be contextual. While the information disclosures our explanations highlight are ones that people care about~\cite{CCS:CumKapRed21}, 
they represent a starting point which should be used to further adapt explanations for specific contexts. The information disclosures we highlight may not be comprehensive of all specific disclosures people are concerned about across contexts. For example, privacy concerns in a particular educational setting may differ from a medical setting. Future work should also study ways to supplement privacy labels for DP with contextually-appropriate communication about the choice of privacy budget~\cite{nanayakkara2023chances,benthall2022}.

\paragraph{Privacy Labels for Other PETs.}
Our approach to designing privacy labels for DP could be adapted to other PETs. We hypothesize that privacy labels that take a contextual integrity approach---emphasizing which data flows are permitted and which are prohibited---could lead to improved comprehension of a variety of PETs~\cite{nissenbaum2009privacy}. Our survey respondents found it more difficult to reason about the implications of local DP than central DP. This finding suggests that clearly explaining which data flows are permitted is particularly important for PETs that enable outsourced computation, such as local DP. Future work could confirm whether the techniques employed here, and the greater difficulty with mental model formation among participants, extends to other PETs that engage in outsourced computation, such as secure multi-party computation, trusted execution environments, and homomorphic encryption

Our findings suggest that people employ their known models of PETs (e.g., understandings of encryption) to reason about new PETs. A standardized approach for presenting the kinds of protection a particular PET offers could help people compare new PETs with more familiar ones. Leveraging this kind of prior knowledge could be beneficial; however, we also caution that in some cases, drawing on knowledge of other PETs could lead to confusion or overtrust. It is important that comparisons between PETs clearly explain their differences and do not overstate the protection offered.

Finally, PETs are rarely deployed in isolation. Our qualitative data show that people are interested in learning about DP \textit{in context}. That is, they want information about the protection offered by DP, but they also care about the other safeguards and signals of trustworthiness that might help them make better-informed holistic data-sharing decisions. Particularly in the case of the central model, users may feel more comfortable if information about DP is presented alongside information about other PETs used to secure user data. Future work should go beyond explaining PETs one at a time and study effective ways to explain the nature of the protection obtained through combinations of PETs. Since our privacy labels focus on information flows---rather than the details of how DP works---it should be straightforward to modify them to communicate the protection offered by multiple PETs in combination.

%% file: 7_appendix.tex
\section{Codes}
\label{app:codes}

Based on the interview data, the research team developed this set of twenty low-level codes, grouped into higher-order themes: \\

\noindent Additional Information Requested
\begin{itemize}[leftmargin=*]
    \item How questions\\
    Example: \textit{Just like how the barrier works, a little more detail.}
    \item What questions\\
    Example: \textit{I would like to know exactly what information from medical records would be shown.}
    \item Who questions\\
    Example: \textit{I don't know what the who the nonprofit is partnering with or why.}
\end{itemize}
Design Feedback
\begin{itemize}[leftmargin=*]
    \item Alternative presentations
    \begin{itemize}
        \item Links to more detailed information\\
        Example: \textit{I'd have a link or something to explain what general patterns means, what's the full detail, maybe as a side if they really were interested in knowing.}
        \item Terms of use / consent documents\\
        Example: \textit{I think I would definitely start with the thing that comes to mind first are informed consents that we sign as participants, and they're very clear about how will your data will be stored and who has access, how will it be de-identified.}
        \item Video or animation\\
        Example: \textit{What you could do is some sort of like animation type thing with a video-like format.}
    \end{itemize}
    \item Icons
    \begin{itemize}
        \item Color\\
        Example: \textit{The green and red doesn't work for me.}
        \item Locks\\
        Example: \textit{I like the look of the lock.}
        \item Privacy barrier\\
        Example: \textit{A label of some sort beyond privacy barrier might be helpful.}
    \end{itemize}
    \item Things people liked\\
    Example: \textit{I like things that make it faster to read.}
\end{itemize}
Participant Understanding
\begin{itemize}[leftmargin=*]
        \item Did not understand\\
        Example: \textit{So I'm not clear as to what the protection actually does.}
        \item Misconception\\
        Example: \textit{It allows the people who utilize the information, say the law enforcement and medical professionals, it would allow them to share that information amongst themselves in a secure network without allowing the people who want to get that information to abuse that information.}
        \item DP as anonymization\\
        Example: \textit{The only way I could explain it would be that an individual's personally identifying details would not be included with their medical records.}
        \item DP as fake data\\
        Example: \textit{ It's basically saying that we might put fake data in some parts of it.}
        \item Other PETs\\
        Example: \textit{So basically there's like  some kind of firewall that keeps my privacy safe.}
        \item User-generated metaphors\\
        Example: \textit{It's kind of like an egg. You know, you crack it open and you don't know if it's going to be rotten inside or not. But I don't know what chicken it came from, so I can't blame the chicken.}
    \end{itemize}
Reasoning About Data Sharing
\begin{itemize}[leftmargin=*]
        \item Benefits\\
        Example: \textit{I actually think that people like data analysts or employee university employees probably want to see my information. Like in that case, that's when it's okay for privacy to be breached. Because it's for the purpose of the study.}
        \item Concerns
        \begin{itemize}
        \item Concerns or skepticism about adequacy of protection\\
        Example: \textit{It sounds good, but I just read too many things about the Internet not being so secure as we would like.}
        \item Data disclosure risks (or lack thereof)\\
        Example: \textit{Especially like insurance companies, I would want to make sure that it's not being shared without my knowledge.}
        \item Lack of concern about privacy in general\\
        Example: \textit{ I don't care about my personal information being released.}
    \end{itemize}
    \end{itemize}

Sets of codes were also developed for the open-text survey responses. The following codes are related to respondents' reasoning about data sharing.

\begin{itemize}[leftmargin=*]
    \item Relationship with doctor\\
    Example: \textit{I believe that if the doctors office is working with the non profit, I believe I trust them, there would also be massive repercussions if they were to do anything wrong with the records.}
    \item Want more info\\
    Example: \textit{before i say yes, i would need more info such as-will they see my name, do they want my entire medical history, what kind of boundaries in medicine are they pushing and do they align with my beliefs and morals}
    \item Too risky or too private\\
    Example: \textit{I think with all that's been going with abortion in the USA I'd be extremely wary of sharing medical data with a third party. Even if they have an extra layer of privacy protection they could still get hacked or the government could decide it has a right to that data.}
    \item Nothing to hide\\
    Example: \textit{I would share my medical records with anyone who wanted to see them. This would not be an issue for me. I have nothing to hide.}
    \item Benefits of data sharing\\
    Example: \textit{Yes, yes and absolutely yes. If this will help just ONE person who needs it, I would gladly share what I can to help them as long as my privacy was protected. Heck, even if it wasn't protected if it could still help then yes. I'm seeing commercials talking about wanting cancer Institutions to start doing this. This could have helped my dad perhaps. And if anyone would need to see his records to help others, I'd say yes.}
    \item Trust\\
    Example: \textit{I want to help them with their research and I trust that they will be able to keep my information private.}
    \item Distrust\\
    Example: \textit{i dont trust them}
    \item Money\\
    Example: \textit{In todays society where information is money, I have a hard time trusting organizations or institutions with very private information such as medical records.}
    \item Frequency of data breaches\\
    Example: \textit{Reassurances about security technology are hollow. Everything is breached eventually. It's just an arms race with the hackers.}
    \item Laws and Regulations\\
    Example: \textit{Medical record information should be protected and private. That is what HIPPA is for.}
    \item Deletion\\
    Example: \textit{Too loose in management, no note of when data will be deleted (which is the basic requirement for data collection in modern times), no mentions of security measures, no compensation for doing so nor any statement on how reputable the nonprofit organization is.}
    \item Data already out there\\
    Example: \textit{Because most medical information is public}
\end{itemize}

This final set of codes is related to feedback obtained through the online survey.
\begin{itemize}[leftmargin=*]
    \item Simplify\\
    Example: \textit{Make it more simplified and shorter}
    \item More info about how protection works\\
    Example: \textit{There needs to be more explanation about how the privacy protection works.}
    \item Other info requests\\
    Example: \textit{need to be informed on where my data is going.}
    \item Positives\\
    Example: \textit{I found the explanations of the privacy protection to be clear, concise, and easy to understand.}
    \item Confusion\\
    Example: \textit{The picture is confusing to me. I don't understand why it needs two different sections}
    \item Nothing is foolproof\\
    Example: \textit{Anything can be hacked. No one can be trusted}
    \item Distrust\\
    Example: \textit{For me it’s more of a feeling that I don’t trust what is being presented as far as the safety of my information.}
\end{itemize}

\section{Survey Instrument}
\label{app:survey}
\subsection{Instructions}
In this survey we are going to ask you a series of questions about a hypothetical scenario. Please do your best to imagine yourself in this scenario and answer the questions as if you were actually making the decisions about which you will be asked.
\subsection{Scenario Description}
Imagine that during your next doctor’s visit, your primary care doctor informs you that they are part of a non-profit organization trying to push the boundaries of medical research. The non-profit is asking patients around the country to share their medical records, which will be used to help medical research on improving treatment options and patient care. Your doctor, with your permission, can facilitate the non-profit getting the information they need. 
\subsection{Privacy Description}
The non-profit organization will use an extra layer of privacy technology to protect your information. [Explanation inserted here.]
\subsection{Comprehension Check}
What kind of information does the non-profit want to collect? [Choice order randomized.]
\begin{itemize}
    \item Medical records
    \item Music videos
    \item Book titles
    \item Location histories
\end{itemize}
\subsection{Trust}
Please indicate your agreement with the following statement:
I trust the non-profit organization to protect my personal information privacy.
\begin{itemize}
    \item Strongly agree
    \item Somewhat agree
    \item Neither agree nor disagree
    \item Somewhat disagree
    \item Strongly disagree
    \item Prefer not to answer
\end{itemize}
\subsection{Self-Efficacy}
How confident are you that you have enough information to decide whether to share your medical record with the non-profit?
\begin{itemize}
    \item Very confident
    \item Confident
    \item Moderately confident
    \item Slightly confident
    \item Not at all confident
    \item Prefer not to answer
\end{itemize}

\noindent
How confident are you about deciding whether to share your medical record with the non-profit?
\begin{itemize}
    \item Very confident
    \item Confident
    \item Moderately confident
    \item Slightly confident
    \item Not at all confident
    \item Prefer not to answer
\end{itemize}

\subsection{Share}
Would you be willing to share your medical record with the non-profit?
\begin{itemize}
    \item Yes
    \item No
    \item Prefer not to answer
\end{itemize}

\noindent
Please explain your decision. [Text entry.]

\subsection{Objective Comprehension}
For each of the following statements, please indicate if you expect the following to be true or false if you share your medical record with the non-profit.

\noindent
An employee working for the non-profit, such as a data analyst, could be able to see my exact medical history.
\begin{itemize}
    \item True
    \item False
    \item I don't know
    \item Prefer not to answer
\end{itemize}

\noindent
A criminal or foreign government that hacks the non-profit could learn my medical history.
\begin{itemize}
    \item True
    \item False
    \item I don't know
    \item Prefer not to answer
\end{itemize}

\noindent
A law enforcement organization could access my medical history with a court order requesting this data from the non-profit.
\begin{itemize}
    \item True
    \item False
    \item I don't know
    \item Prefer not to answer
\end{itemize}

\noindent
Graphs or informational charts created using information given to the non-profit could reveal my medical history.
\begin{itemize}
    \item True
    \item False
    \item I don't know
    \item Prefer not to answer
\end{itemize}

\noindent
Data that the non-profit shares with other organizations doing medical research could reveal my medical history.
\begin{itemize}
    \item True
    \item False
    \item I don't know
    \item Prefer not to answer
\end{itemize}

\subsection{Thoroughness}
Please indicate your agreement with the following statement:
I feel that it was explained thoroughly to me how the non-profit protects patient privacy.
\begin{itemize}
    \item Strongly agree
    \item Somewhat agree
    \item Neither agree nor disagree
    \item Somewhat disagree
    \item Strongly disagree
    \item Prefer not to answer
\end{itemize}

\subsection{Subjective Understanding}
How confident are you in your understanding of the privacy protection?
\begin{itemize}
    \item Very confident
    \item Confident
    \item Moderately confident
    \item Slightly confident
    \item Not at all confident
    \item Prefer not to answer
\end{itemize}

\subsection{Feedback}
What feedback (if any) would you like to share about the explanations of privacy protection? [Text entry.]

\subsection{PETs}
Have you ever heard of the following technologies? (select all that apply) [Choice order randomized.]
\begin{itemize}
    \item Differential privacy
    \item End-to-end encryption
    \item Secure multi-party computation
    \item Deliquescent security
    \item None of the above
    \item Prefer not to answer
\end{itemize}

\noindent
Which of these technologies do you think was described in the survey? [Choice order randomized.]
\begin{itemize}
    \item Differential privacy
    \item End-to-end encryption
    \item Secure multi-party computation
    \item Deliquescent security
    \item None of the above
    \item Prefer not to answer
\end{itemize}

\noindent
Please explain your reasoning. [Text entry.]

\subsection{Background}
How familiar are you with the following computer and Internet-related items? 
\begin{itemize}
    \item Advanced Search \\1 (No Understanding) - 5 (Full understanding)\\ or Prefer not to answer
    \item PDF 1 (No Understanding) - 5 (Full understanding)\\ or Prefer not to answer
    \item Spyware\\ 1 (No Understanding) - 5 (Full understanding)\\ or Prefer not to answer
    \item Wiki\\ 1 (No Understanding) - 5 (Full understanding)\\ or Prefer not to answer
    \item Cache\\ 1 (No Understanding) - 5 (Full understanding)\\ or Prefer not to answer
    \item Phishing\\ 1 (No Understanding) - 5 (Full understanding)\\ or Prefer not to answer
\end{itemize}

\noindent
In what year were you born? (four digits please) [Text entry.]

\noindent
What is your gender? [Multiselect.]
\begin{itemize}
    \item Man
    \item Woman
    \item Non-binary
    \item Prefer to self describe: [Text entry.]
    \item Prefer not to answer
\end{itemize}

\noindent
Please specify your race/ethnicity (select all that apply).
\begin{itemize}
    \item Hispanic, Latino, or Spanish
    \item Black or African American
    \item White
    \item American Indian or Alaska Native
    \item Asian, Native Hawaiian, or Pacific Islander
    \item Prefer to self describe: [Text entry.]
    \item Prefer not to answer
\end{itemize}

\noindent
What is the highest level of school you have completed or the highest degree you have received?
\begin{itemize}
    \item Less than high school degree
    \item High school graduate (high school diploma or equivalent including GED)
    \item Some college but no degree
    \item Associate's degree
    \item Bachelor's degree
    \item Advanced degree (e.g., Master's, doctorate)
    \item Prefer not to answer
\end{itemize}

\noindent
Which of the following best describes your educational background or job field?
\begin{itemize}
    \item I have an education in, or work in, the field of computer science, computer engineering or IT.
    \item I DO NOT have an education in, nor do I work in, the field of computer science, computer engineering or IT.
    \item Prefer not to answer
\end{itemize}

\noindent
Which one of the following includes your total HOUSEHOLD income for last year, before taxes?
\begin{itemize}
    \item Less than \$10,000
    \item \$10,000 to under \$20,000
    \item \$20,000 to under \$30,000
    \item \$30,000 to under \$40,000
    \item \$40,000 to under \$50,000
    \item \$50,000 to under \$65,000
    \item \$65,000 to under \$80,000
    \item \$80,000 to under \$100,000
    \item \$100,000 to under \$125,000
    \item \$125,000 to under \$150,000
    \item \$150,000 to under \$200,000
    \item \$200,000 or more
    \item Prefer not to answer
\end{itemize}

\section{Demographics}
\label{app:demographics}
Table~\ref{tab:demographics1} describes the demographics of the 24 participants in the main interview study. Table~\ref{tab:demographics3} describes the demographics of the 10 participants who participated in the follow-up interviews. Table~\ref{tab:demographics2} summarizes the demographics of the survey respondents. Note that respondents could select multiple values for race/ethnicity and gender and that many respondents selected multiple options for race/ethnicity but did not explicitly describe themselves as multiracial. Table~\ref{tab:pet} displays the approximate percentage of respondents who expressed familiarity with various PETs out of all respondents who answered this question (n=684).

\begin{minipage}[t]{0.49\textwidth}%
              \centering
              \scriptsize
                    \captionof{table}{Participant Demographics: Initial Interviews}
                    \begin{tabular}{l l r}
                    \toprule
                    \multicolumn{2}{l}{\textbf{Demographic Attribute}} &  \textbf{Count} \\
                    \midrule
                    \textbf{\emph{Gender}} & Female & 10 \\
                    & Male & 14 \\
                    \hline
                    \textbf{\emph{Age}} & < 20 & 2 \\
                     & 20-29 & 9 \\
                     & 30-39 & 6 \\
                     & 40-49 & 4 \\
                     & 50+ & 3 \\
                    \hline
                    \textbf{\emph{Race}} & Asian & 1 \\
                     & Black or African American & 4 \\
                     & Mixed, Multiracial, or Biracial & 3 \\
                     & White or Caucasian & 16 \\
                    \hline
                    \textbf{\emph{Education}} & Secondary education (e.g. GED / GCSE) & 1 \\
                     & High school diploma / A-levels & 11 \\
                     & Technical / community college & 4 \\
                     & Undergraduate degree (BA / BSc / other) & 5 \\
                     & Graduate degree & 2 \\
                     & Doctorate degree (PhD / other) & 1 \\
                    \bottomrule
                    \end{tabular}
                    \label{tab:demographics1}
    \end{minipage}%
    \begin{minipage}[t]{0.49\textwidth}%
                \centering
                \scriptsize
                \captionof{table}{Participant Demographics: Follow-up Interviews}
                \begin{tabular}{l l r}
                \toprule
                \multicolumn{2}{l}{\textbf{Demographic Attribute}} & \textbf{Count} \\
                \midrule
                \textbf{\emph{Gender}} & Female & 5 \\
                 & Male & 5 \\
                \hline
                \textbf{\emph{Age}} & < 20 & 1 \\
                 & 20-29 & 3 \\
                 & 30-39 & 1 \\
                 & 40-49 & 1 \\
                 & 50+ & 4 \\
                \hline
                \textbf{\emph{Race}} & Asian & 3 \\
                 & Black or African American & 1 \\
                 & Mixed, Multiracial, or Biracial & 2 \\
                 & White or Caucasian & 3 \\
                 & Native American & 1 \\
                \hline
                \textbf{\emph{Education}} & High school diploma / A-levels & 2 \\
                 & Technical / community college & 2 \\
                 & Undergraduate degree (BA / BSc / other) & 5 \\
                 & Doctorate degree (PhD / other) & 1 \\
                \bottomrule
                \end{tabular}
                \label{tab:demographics3}
    \end{minipage}\\\\

\begin{minipage}[t]{0.59\textwidth}%
      \centering
      \scriptsize
    \captionof{table}{Respondent Demographics}
    \begin{tabular}{l l r}
    \toprule
    \multicolumn{2}{l}{\textbf{Demographic Attribute}} & \textbf{Count} \\
    \midrule
    \textbf{\emph{Gender}} & Woman & 343 \\
     & Man & 335 \\
     & Non-binary & 15 \\
     & Agender / Gender-fluid afab / genderqueer / they & 5 \\
    \hline
    \textbf{\emph{Age}} & < 20 & 12 \\
     & 20-29 & 249 \\
     & 30-39 & 219 \\
     & 40-49 & 98 \\
     & 50+ & 119 \\
    \hline
    \textbf{\emph{Race/Ethnicity}} & Hispanic, Latino, or Spanish & 83 \\
     & Black or African American & 68 \\
     & White & 478 \\
     & American Indian or Alaska Native & 12 \\
     & Asian, Native Hawaiian, or Pacific Islander & 110 \\
     & Multiracial or Mixed race & 4 \\
    \hline
    \textbf{\emph{Education}} & High school or less & 124 \\
     & Some college &  233 \\
     & Bachelor’s or above & 337 \\
    \hline
    \textbf{\emph{Income}} & Less than \$10,000 & 41\\
     & \$10,000 to under \$20,000 & 53\\
     & \$20,000 to under \$30,000 & 79\\
     & \$30,000 to under \$40,000 & 68\\
     & \$40,000 to under \$50,000 &  65\\
     & \$50,000 to under \$65,000 & 85\\
     & \$65,000 to under \$80,000 & 88\\
     & \$80,000 to under \$100,000 & 51\\
     & \$100,000 to under \$125,000 & 57\\
     & \$125,000 to under \$150,000 & 30\\
     & \$150,000 to under \$200,000 & 25 \\
     & \$200,000 or more & 32\\
    \hline
    \textbf{\emph{Tech}} & Education or work in CSE/IT & 148 \\
     & No education nor work in CSE/IT &  527 \\
    \bottomrule
    \end{tabular}
    \label{tab:demographics2}
\end{minipage}%
\begin{minipage}[t]{0.4\textwidth}%
    \centering
    \scriptsize
    \captionof{table}{Familiarity with PETs}
    \begin{tabular}{lll}
    \toprule
    PET&\#&\%\\
    \midrule
    {\small End-to-end encryption} & {\small 439} & {\small 64\%}\\
    {\small Differential privacy} & {\small 32} & {\small 5\%}\\
    {\small Secure multi-party computation} & {\small 26} & {\small 4\%}\\
    {\small Deliquescent security (distractor)} & {\small 3} & {\small <1\%}\\
    {\small None of the above} & {\small 237} & {\small 35\%} \\
    \bottomrule
    \end{tabular}
    \label{tab:pet}
\end{minipage}%

\newpage
\section{Designs} 
\label{app:designs}
\begin{figure}[!htbp]
\centering
\begin{subfigure}{.7\textwidth}
  \centering
    \includegraphics[width=.7\linewidth]{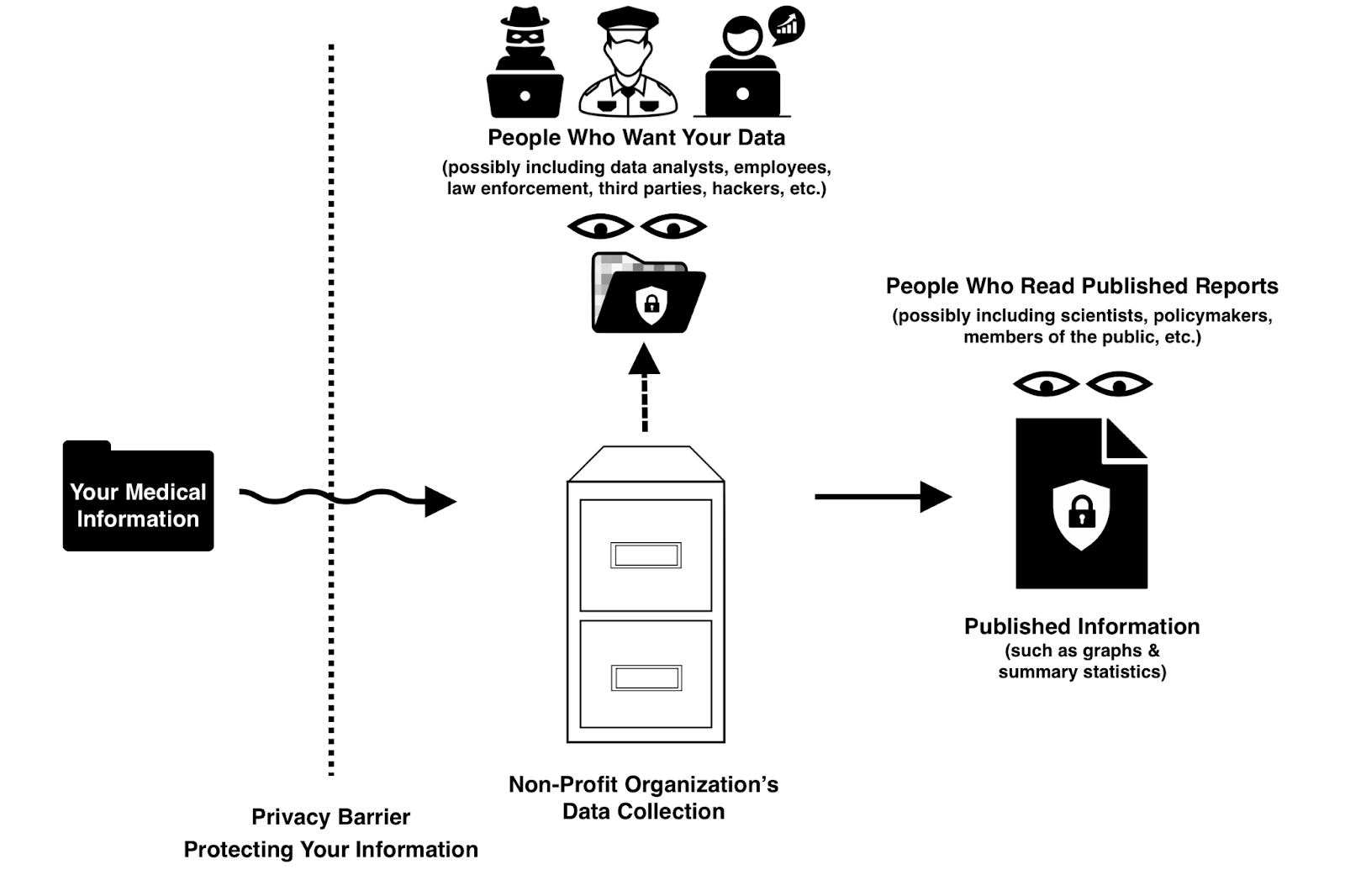}
  \label{fig:diagramslocal}
\end{subfigure}
\par\bigskip
\par\bigskip
\begin{subfigure}{.7\textwidth}
\centering
    \includegraphics[width=.7\linewidth]{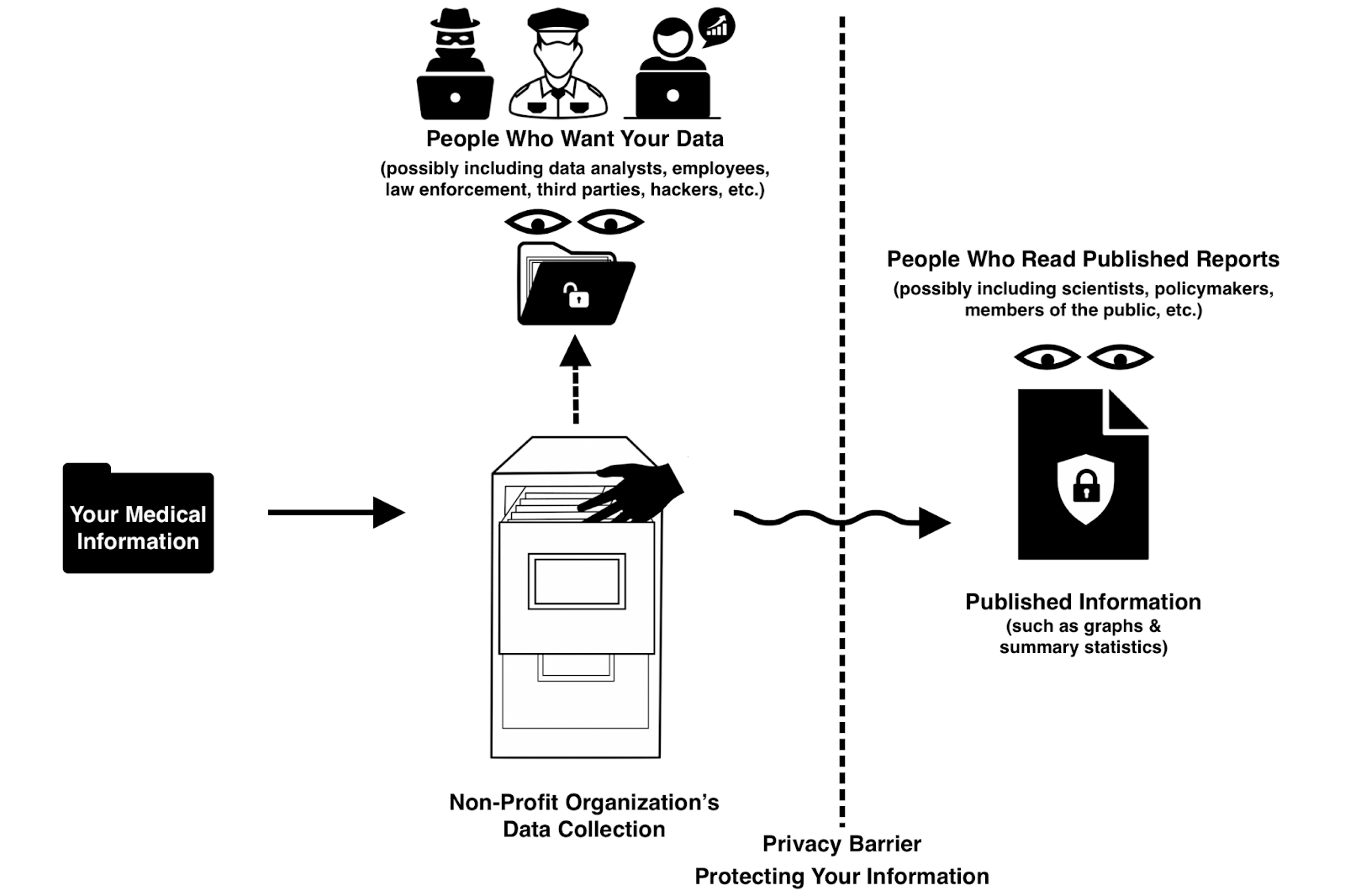}
  \label{fig:diagramscentral}
\end{subfigure}
\caption{Top: Diagram for local model. Bottom: Diagram for central model.}
\label{fig:diagrams}
\end{figure}

\begin{figure*}[!htbp]
\centering
\begin{subfigure}[T]{.5\textwidth}
  \centering
  \includegraphics[width=.95\linewidth]{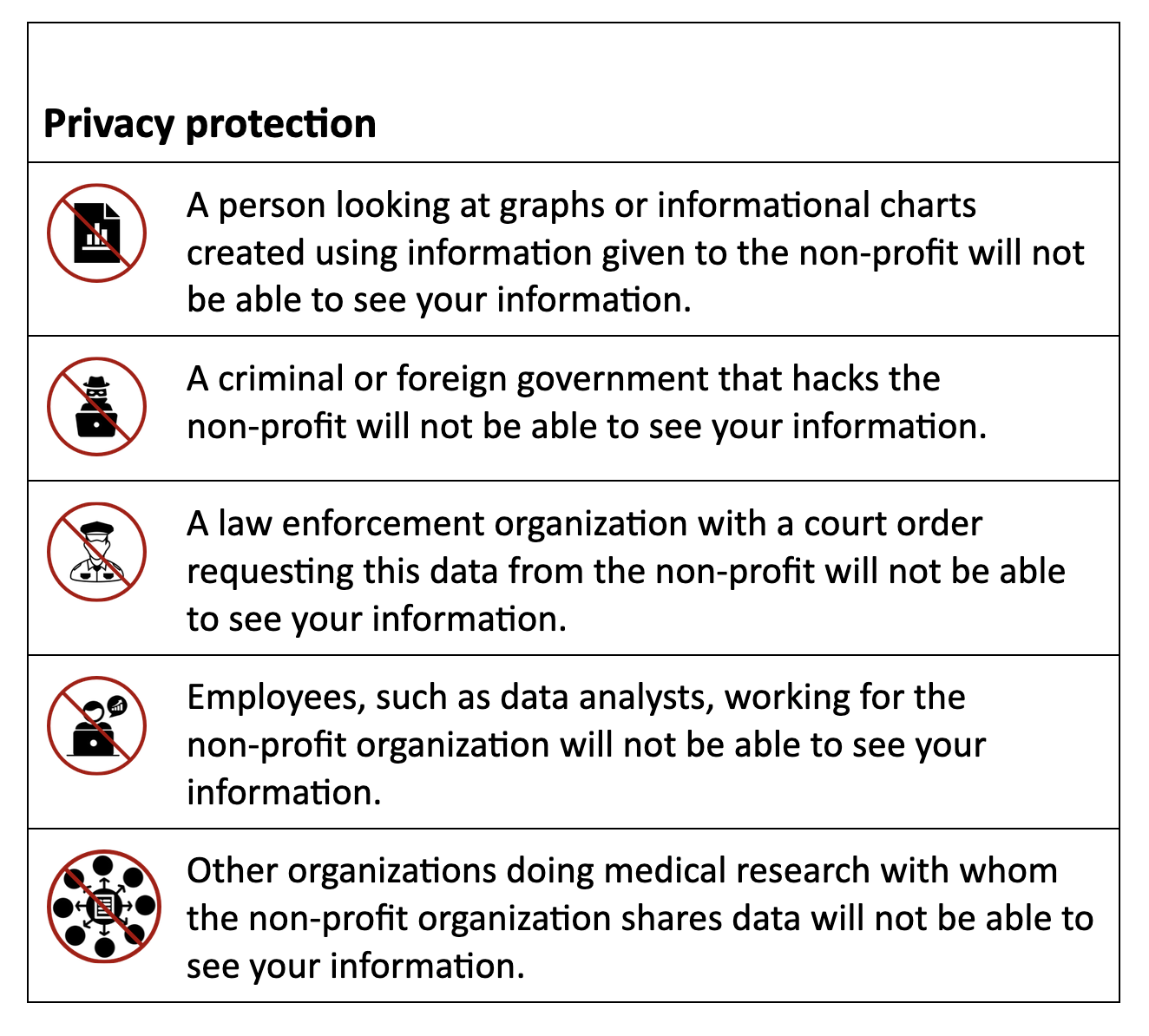}
  \label{fig:tablocalv1}
\end{subfigure}%
\begin{subfigure}[T]{.5\textwidth}
  \centering
  \includegraphics[width=.95\linewidth]{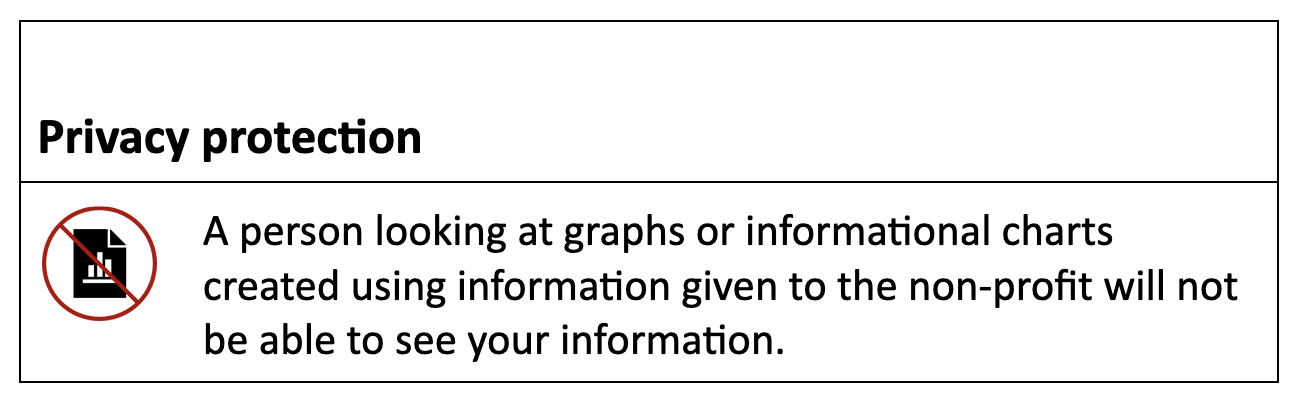}
  \label{fig:tabcentralv1}
\end{subfigure}
\begin{subfigure}[t]{.5\textwidth}
  \centering
  \includegraphics[width=.95\linewidth]{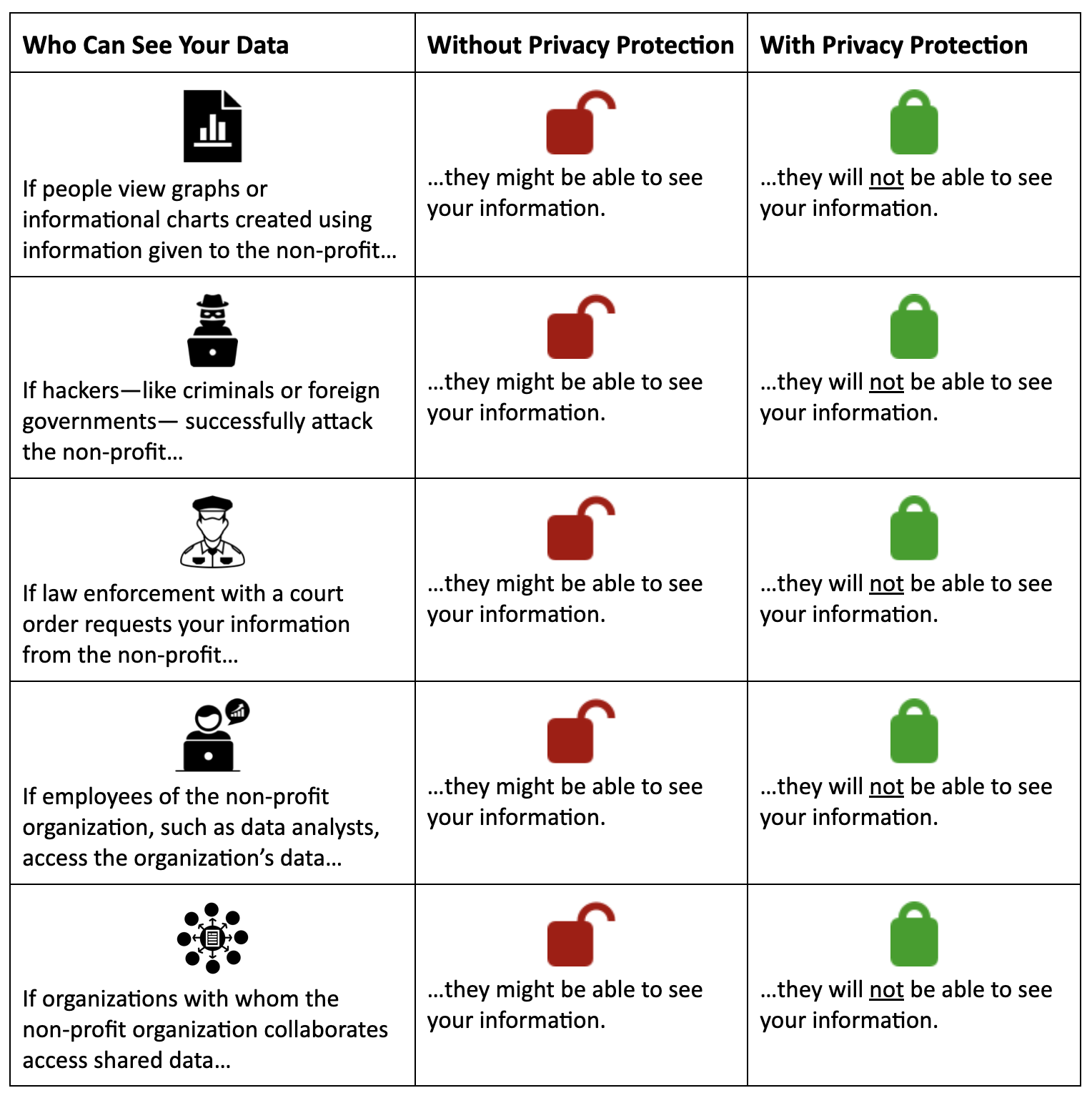}
  \caption{Local}
  \label{fig:tablocalv2}
\end{subfigure}%
\begin{subfigure}[t]{.5\textwidth}
  \centering
  \includegraphics[width=.95\linewidth]{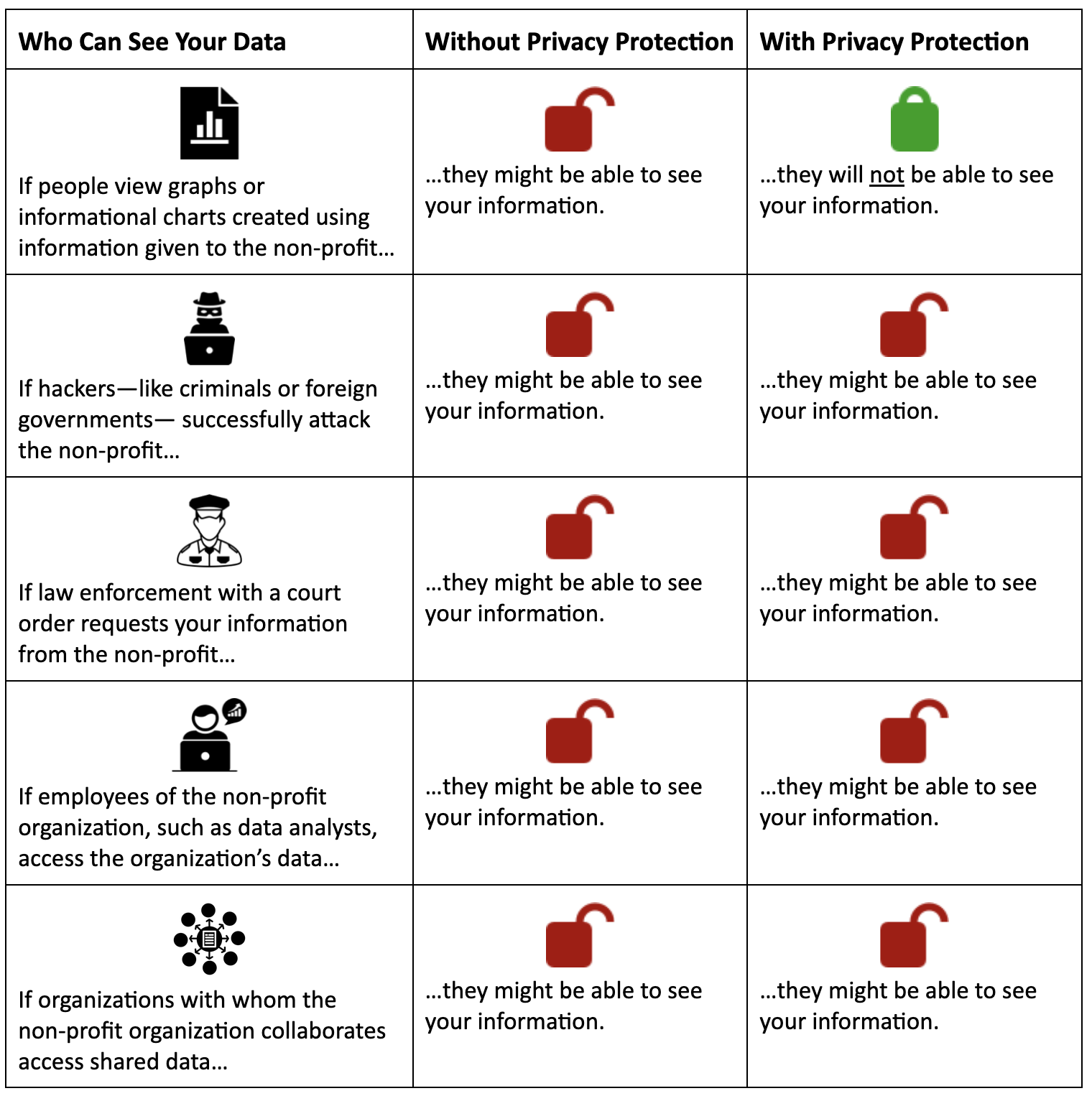}
  \caption{Central}
  \label{fig:tabcentralv2}
\end{subfigure}
\caption{Original Privacy Labels.}
\label{fig:tab}
\end{figure*}

\begin{figure*}[!htbp]
\centering

  \includegraphics[width=.95\linewidth]{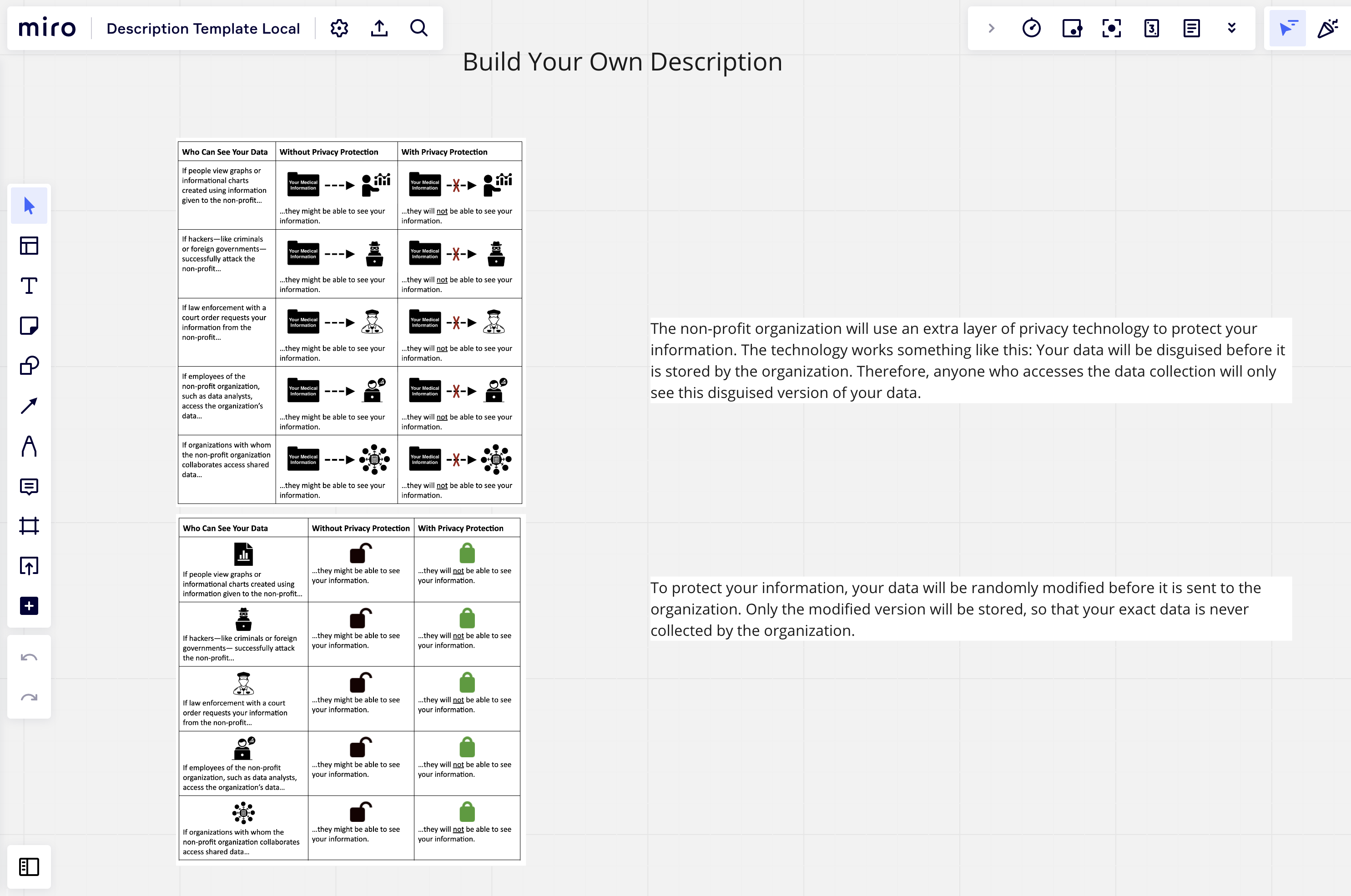}

\caption{Example of the Miro board setup used for the follow-up interviews.}
\label{fig:miro}
\end{figure*}

\begin{figure*}[!htbp]
\centering

  \includegraphics[width=.99\linewidth]{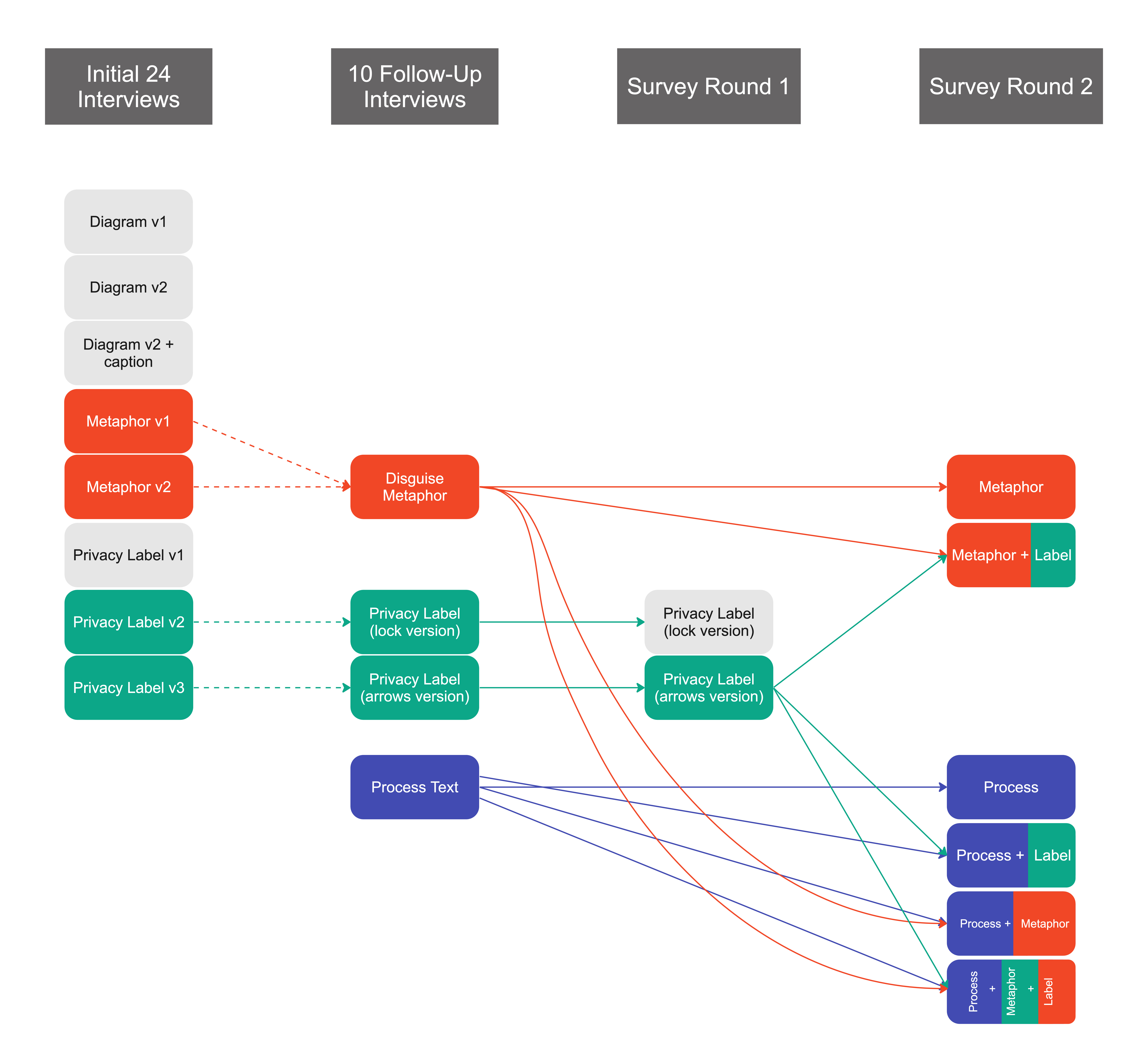}

\caption{In the initial interviews, we evaluated multiple versions of each explanation type for both the local and central models. Based on participant feedback, we dropped the diagram explanations and modified the privacy labels. During the follow-up interviews, we developed a new metaphor and introduced a text with information about the data protection process. Next, we compared the two privacy labels through a survey, and dropped the version with locks. Finally, we evaluated the disguise metaphor, the process text, and combinations of these texts and the privacy label with arrows.}
\label{fig:flow}
\end{figure*}

Table~\ref{tab:metaphors} lists all of the original metaphor texts. Figure~\ref{fig:diagrams} shows representative examples of our diagrams, and figure~\ref{fig:tab} shows our privacy labels. Figure~\ref{fig:miro} shows an example of the kind of Miro board that a participant in one of our follow-up interviews would have interacted with. Figure~\ref{fig:flow} shows how our designs evolved over time. Table~\ref{tab:allconditions} lists all survey conditions. 

\input{tables/metaphors}

\input{tables/allexplanations}

\section{Descriptive Statistics}
\label{app:stat}
Table~\ref{tab:descr} displays the proportion of respondents per condition who answered each comprehension correctly ({\color{DarkGreen}{+}}) and incorrectly ({\color{DarkRed}{-}}). Since some respondents selected `I don't know,` these percentages may not add to 1. 

\input{tables/descriptive}

%% file: tables/metaphors.tex
\begin{table}[!htbp]
  \centering
  \caption{Original Metaphor Descriptions}
  \begin{tabular}{p{0.43\linewidth} p{0.49\linewidth}}
    \toprule
    \multicolumn{1}{c}{\textbf{Local}} & \multicolumn{1}{c}{\textbf{Central}} \\
    \toprule
    {\small \textit{Sharing data with the protection of this technology is like donating a penny to a crowdfunding campaign. No one will know with certainty that you donated. The sum of the donations from a large group of people will be valuable to our data analysts.}} 
    & 
    {\small \textit{Publishing statistics, graphs, or tables using this technology is like publishing a blurry photo of the database that allows the viewer to see general patterns while hiding individual details. However, someone who obtained access to the database would be able to see all of the collected information in full detail.}} \\
    \midrule
    {\small \textit{The technology works something like this: Imagine that we are collecting photographs, but instead of collecting the raw images, we blur the images, and only collect the blurry images, so that little is revealed about you as an individual. Anyone who accesses our collected data will only see the blurry images, rather than the originals.}} 
    & 
    {\small \textit{Publishing statistics, graphs, or tables using this technology is like publishing a photo of a mosaic, taken from a distance. People viewing this photo would not be able to see the individual tiles\textemdash in other words, individuals’ data\textemdash yet they would still be able to see the overall picture. However, someone with direct access to the mosaic would be able to discern the individual tiles.}}
    \\
    \bottomrule
  \end{tabular}
  \label{tab:metaphors}
\end{table}

%% file: tables/allexplanations.tex
\begin{table*}[!htbp]
{\small
  \caption{All Explanation Texts}
  \label{tab:allconditions}
  \begin{tabular}{l m{7cm} m{7.5cm}}
    \toprule
    Type&Local&Central\\
    \midrule
    {\scriptsize Arrows Label}  & \includegraphics[width=0.38\textwidth]{figures/ArrowsLocal.png} & \includegraphics[width=0.38\textwidth]{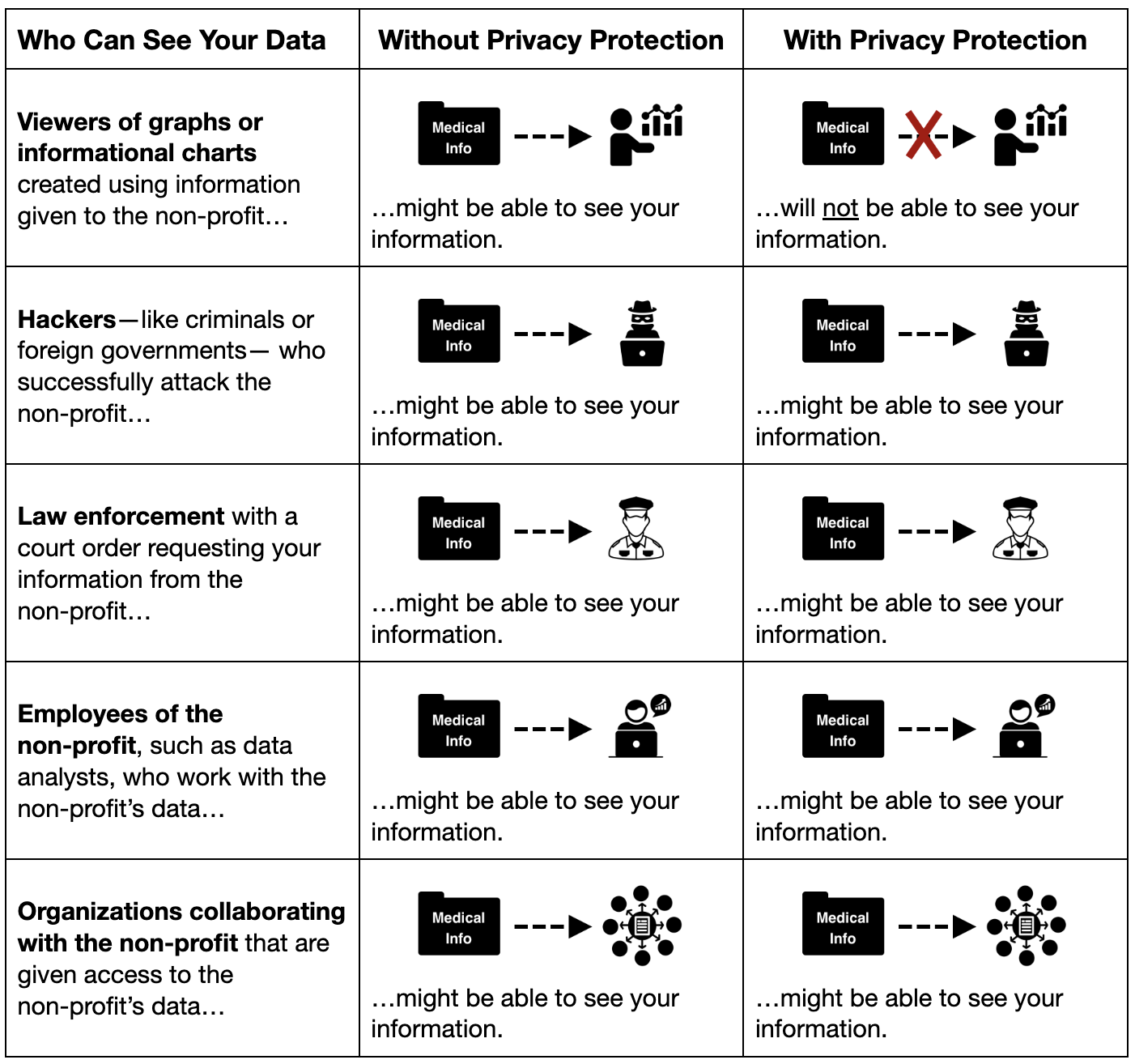} \\
    \midrule
    {\scriptsize Process} & {\scriptsize To protect your information, your data will be randomly modified before it is sent to the organization. Only the modified version will be stored, so that your exact data is never collected by the organization.} & {\scriptsize To protect your information, the organization will store your data but only publish reports, graphs, or charts that have been randomly modified. These modifications hide information that is unique to you as an individual.}\\
    \midrule
   {\scriptsize Metaphor} & {\scriptsize The technology works something like this: Your data will be disguised before it is stored by the organization. Therefore, anyone who accesses the data collection will only see this disguised version of your data.} & {\scriptsize The technology works something like this: The collected data will be disguised when any graphs, charts, or reports are published. However, anyone who accesses the organization’s data collection will see the undisguised data.} \\
   \midrule
   {\scriptsize Metaphor+Process} & {\scriptsize The technology works something like this: Your data will be disguised before it is stored by the organization. Therefore, anyone who accesses the data collection will only see this disguised version of your data. More specifically, your data will be randomly modified before it is sent to the organization. Only the modified version will be stored, so that your exact data is never collected by the organization.} & {\scriptsize The technology works something like this: The collected data will be disguised when any graphs, charts, or reports are published. However, anyone who accesses the organization’s data collection will see the undisguised data. More specifically, the organization will store your data but only publish reports, graphs, or charts that have been randomly modified. These modifications hide information that is unique to you as an individual.} \\
   \midrule
   {\scriptsize Label+Metaphor}  & {\scriptsize See \textit{Arrows Label} and \textit{Metaphor} rows.}  \\
   \midrule
   {\scriptsize Label+Process} & {\scriptsize See \textit{Arrows Label} and \textit{Process} rows.}  \\
   \midrule
   {\scriptsize Label+Process+Metaphor} & \scriptsize See \textit{Arrows Label} and \textit{Metaphor+Process} rows.\\
   \midrule
   {\scriptsize Xiong et al.} & {\scriptsize To respect your personal information privacy and ensure best user experience, the data shared with the non-profit organization will be processed via an additional privacy technique. That is, your data will be randomly modified before it is sent to the organization. Since the organization stores only the modified version of your personal information, your privacy is protected even if the organization's database is compromised.} & { \scriptsize To respect your personal information privacy and ensure best user experience, the data shared with the non-profit organization will be processed via an additional privacy technique. That is, the organization will store your data but only publish the aggregated statistics with modification so that your personal information cannot be learned. However, your personal information may be leaked if the organization’s database is compromised.}\\
  \bottomrule
\end{tabular}
}
\end{table*}

%% file: tables/descriptive.tex
\begin{table}[h]
\centering  
\caption{Accuracy of Privacy Expectations}
  \begin{tabular}{ll|cc|cc|cc|cc|cc}
    \toprule
    Model& Explanation&\multicolumn{2}{c|}{Hack}&\multicolumn{2}{c|}{Law}&\multicolumn{2}{c|}{Org}&\multicolumn{2}{c|}{Graph}&\multicolumn{2}{c}{Share}\\
     & &
     {\scriptsize \color{DarkGreen}{+}}&
     {\scriptsize \color{DarkRed}{-}}&
     {\scriptsize \color{DarkGreen}{+}}&
     {\scriptsize \color{DarkRed}{-}}&
     {\scriptsize \color{DarkGreen}{+}}&
     {\scriptsize \color{DarkRed}{-}}&
     {\scriptsize \color{DarkGreen}{+}}&
     {\scriptsize \color{DarkRed}{-}}&
     {\scriptsize \color{DarkGreen}{+}}&
     {\scriptsize \color{DarkRed}{-}}\\
     
    \midrule

    {\scriptsize Central} & {\scriptsize Metaphor} 
    & {\scriptsize \color{DarkGreen}{0.78}} & {\scriptsize \color{DarkRed}{0.05}} 
    & {\scriptsize \color{DarkGreen}{0.54}} & {\scriptsize \color{DarkRed}{0.14}} 
    & {\scriptsize \color{DarkGreen}{0.89}} & {\scriptsize \color{DarkRed}{0.03}} 
    & {\scriptsize \color{DarkGreen}{0.41}} & {\scriptsize \color{DarkRed}{0.46}} 
    & {\scriptsize \color{DarkGreen}{0.54}} & {\scriptsize \color{DarkRed}{0.22}}\\

    {\scriptsize Local} & {\scriptsize Metaphor} 
    & {\scriptsize \color{DarkGreen}{0.26}} & {\scriptsize \color{DarkRed}{0.47}} 
    & {\scriptsize \color{DarkGreen}{0.24}} & {\scriptsize \color{DarkRed}{0.45}} 
    & {\scriptsize \color{DarkGreen}{0.37}} & {\scriptsize \color{DarkRed}{0.37}} 
    & {\scriptsize \color{DarkGreen}{0.39}} & {\scriptsize \color{DarkRed}{0.42}} 
    & {\scriptsize \color{DarkGreen}{0.26}} & {\scriptsize \color{DarkRed}{0.47}}\\

    {\scriptsize Central} & {\scriptsize Process} 
    & {\scriptsize \color{DarkGreen}{0.50}} & {\scriptsize \color{DarkRed}{0.28}} 
    & {\scriptsize \color{DarkGreen}{0.50}} & {\scriptsize \color{DarkRed}{0.17}} 
    & {\scriptsize \color{DarkGreen}{0.52}} & {\scriptsize \color{DarkRed}{0.17}} 
    & {\scriptsize \color{DarkGreen}{0.42}} & {\scriptsize \color{DarkRed}{0.32}} 
    & {\scriptsize \color{DarkGreen}{0.45}} & {\scriptsize \color{DarkRed}{0.32}}\\

    {\scriptsize Local} & {\scriptsize Process} 
    & {\scriptsize \color{DarkGreen}{0.28}} & {\scriptsize \color{DarkRed}{0.48}} 
    & {\scriptsize \color{DarkGreen}{0.3}} & {\scriptsize \color{DarkRed}{0.50}} 
    & {\scriptsize \color{DarkGreen}{0.38}} & {\scriptsize \color{DarkRed}{0.42}} 
    & {\scriptsize \color{DarkGreen}{0.32}} & {\scriptsize \color{DarkRed}{0.28}} 
    & {\scriptsize \color{DarkGreen}{0.28}} & {\scriptsize \color{DarkRed}{0.52}}\\

    {\scriptsize Central} & {\scriptsize Process+Metaphor} 
    & {\scriptsize \color{DarkGreen}{0.79}} & {\scriptsize \color{DarkRed}{0.11}} 
    & {\scriptsize \color{DarkGreen}{0.76}} & {\scriptsize \color{DarkRed}{0.05}} 
    & {\scriptsize \color{DarkGreen}{0.92}} & {\scriptsize \color{DarkRed}{0.08}} 
    & {\scriptsize \color{DarkGreen}{0.61}} & {\scriptsize \color{DarkRed}{0.34}} 
    & {\scriptsize \color{DarkGreen}{0.50}} & {\scriptsize \color{DarkRed}{0.37}}\\

    {\scriptsize Local} & {\scriptsize Process+Metaphor} 
    & {\scriptsize \color{DarkGreen}{0.36}} & {\scriptsize \color{DarkRed}{0.38}} 
    & {\scriptsize \color{DarkGreen}{0.33}} & {\scriptsize \color{DarkRed}{0.33}} 
    & {\scriptsize \color{DarkGreen}{0.44}} & {\scriptsize \color{DarkRed}{0.41}} 
    & {\scriptsize \color{DarkGreen}{0.49}} & {\scriptsize \color{DarkRed}{0.31}} 
    & {\scriptsize \color{DarkGreen}{0.38}} & {\scriptsize \color{DarkRed}{0.41}}\\

    {\scriptsize Central} & {\scriptsize ArrowLabel} 
    & {\scriptsize \color{DarkGreen}{0.92}} & {\scriptsize \color{DarkRed}{0.03}} 
    & {\scriptsize \color{DarkGreen}{0.95}} & {\scriptsize \color{DarkRed}{0.00}} 
    & {\scriptsize \color{DarkGreen}{0.92}} & {\scriptsize \color{DarkRed}{0.05}} 
    & {\scriptsize \color{DarkGreen}{0.58}} & {\scriptsize \color{DarkRed}{0.26}} 
    & {\scriptsize \color{DarkGreen}{0.82}} & {\scriptsize \color{DarkRed}{0.00}}\\

    {\scriptsize Local} & {\scriptsize ArrowLabel} 
    & {\scriptsize \color{DarkGreen}{0.51}} & {\scriptsize \color{DarkRed}{0.32}} 
    & {\scriptsize \color{DarkGreen}{0.44}} & {\scriptsize \color{DarkRed}{0.37}} 
    & {\scriptsize \color{DarkGreen}{0.46}} & {\scriptsize \color{DarkRed}{0.32}} 
    & {\scriptsize \color{DarkGreen}{0.63}} & {\scriptsize \color{DarkRed}{0.24}} 
    & {\scriptsize \color{DarkGreen}{0.63}} & {\scriptsize \color{DarkRed}{0.27}}\\

    {\scriptsize Central} & {\scriptsize Label+Metaphor} 
    & {\scriptsize \color{DarkGreen}{0.85}} & {\scriptsize \color{DarkRed}{0.13}} 
    & {\scriptsize \color{DarkGreen}{0.82}} & {\scriptsize \color{DarkRed}{0.05}} 
    & {\scriptsize \color{DarkGreen}{0.82}} & {\scriptsize \color{DarkRed}{0.08}} 
    & {\scriptsize \color{DarkGreen}{0.64}} & {\scriptsize \color{DarkRed}{0.26}} 
    & {\scriptsize \color{DarkGreen}{0.77}} & {\scriptsize \color{DarkRed}{0.15}}\\

    {\scriptsize Local} & {\scriptsize Label+Metaphor} 
    & {\scriptsize \color{DarkGreen}{0.72}} & {\scriptsize \color{DarkRed}{0.22}} 
    & {\scriptsize \color{DarkGreen}{0.58}} & {\scriptsize \color{DarkRed}{0.33}} 
    & {\scriptsize \color{DarkGreen}{0.58}} & {\scriptsize \color{DarkRed}{0.33}} 
    & {\scriptsize \color{DarkGreen}{0.69}} & {\scriptsize \color{DarkRed}{0.19}} 
    & {\scriptsize \color{DarkGreen}{0.61}} & {\scriptsize \color{DarkRed}{0.31}}\\

    {\scriptsize Central} & {\scriptsize Label+Process} 
    & {\scriptsize \color{DarkGreen}{0.85}} & {\scriptsize \color{DarkRed}{0.12}} 
    & {\scriptsize \color{DarkGreen}{0.82}} & {\scriptsize \color{DarkRed}{0.05}} 
    & {\scriptsize \color{DarkGreen}{0.65}} & {\scriptsize \color{DarkRed}{0.15}} 
    & {\scriptsize \color{DarkGreen}{0.6}} & {\scriptsize \color{DarkRed}{0.25}} 
    & {\scriptsize \color{DarkGreen}{0.62}} & {\scriptsize \color{DarkRed}{0.22}}\\

    {\scriptsize Local} & {\scriptsize Label+Process} 
    & {\scriptsize \color{DarkGreen}{0.59}} & {\scriptsize \color{DarkRed}{0.28}} 
    & {\scriptsize \color{DarkGreen}{0.44}} & {\scriptsize \color{DarkRed}{0.31}} 
    & {\scriptsize \color{DarkGreen}{0.67}} & {\scriptsize \color{DarkRed}{0.21}} 
    & {\scriptsize \color{DarkGreen}{0.62}} & {\scriptsize \color{DarkRed}{0.18}} 
    & {\scriptsize \color{DarkGreen}{0.67}} & {\scriptsize \color{DarkRed}{0.28}}\\

    {\scriptsize Central} & {\scriptsize Label+Process+Metaphor} 
    & {\scriptsize \color{DarkGreen}{0.85}} & {\scriptsize \color{DarkRed}{0.05}} 
    & {\scriptsize \color{DarkGreen}{0.82}} & {\scriptsize \color{DarkRed}{0.03}} 
    & {\scriptsize \color{DarkGreen}{0.80}} & {\scriptsize \color{DarkRed}{0.15}} 
    & {\scriptsize \color{DarkGreen}{0.45}} & {\scriptsize \color{DarkRed}{0.40}} 
    & {\scriptsize \color{DarkGreen}{0.70}} & {\scriptsize \color{DarkRed}{0.15}}\\

    {\scriptsize Local} & {\scriptsize Label+Process+Metaphor} 
    & {\scriptsize \color{DarkGreen}{0.56}} & {\scriptsize \color{DarkRed}{0.26}} 
    & {\scriptsize \color{DarkGreen}{0.59}} & {\scriptsize \color{DarkRed}{0.18}} 
    & {\scriptsize \color{DarkGreen}{0.56}} & {\scriptsize \color{DarkRed}{0.26}} 
    & {\scriptsize \color{DarkGreen}{0.69}} & {\scriptsize \color{DarkRed}{0.15}} 
    & {\scriptsize \color{DarkGreen}{0.49}} & {\scriptsize \color{DarkRed}{0.26}}\\

    {\scriptsize Central} & {\scriptsize Xiong} 
    & {\scriptsize \color{DarkGreen}{0.91}} & {\scriptsize \color{DarkRed}{0.03}} 
    & {\scriptsize \color{DarkGreen}{0.44}} & {\scriptsize \color{DarkRed}{0.12}} 
    & {\scriptsize \color{DarkGreen}{0.62}} & {\scriptsize \color{DarkRed}{0.15}} 
    & {\scriptsize \color{DarkGreen}{0.35}} & {\scriptsize \color{DarkRed}{0.44}} 
    & {\scriptsize \color{DarkGreen}{0.44}} & {\scriptsize \color{DarkRed}{0.32}}\\

    {\scriptsize Local} & {\scriptsize Xiong} 
    & {\scriptsize \color{DarkGreen}{0.33}} & {\scriptsize \color{DarkRed}{0.31}} 
    & {\scriptsize \color{DarkGreen}{0.23}} & {\scriptsize \color{DarkRed}{0.44}} 
    & {\scriptsize \color{DarkGreen}{0.33}} & {\scriptsize \color{DarkRed}{0.41}} 
    & {\scriptsize \color{DarkGreen}{0.56}} & {\scriptsize \color{DarkRed}{0.23}} 
    & {\scriptsize \color{DarkGreen}{0.41}} & {\scriptsize \color{DarkRed}{0.46}}\\

  \bottomrule
\end{tabular}
\label{tab:descr}
\end{table}

%% file: main.bbl
\begin{thebibliography}{100}

\bibitem{abdul2020cogam}
{\sc Abdul, A., von~der Weth, C., Kankanhalli, M., and Lim, B.~Y.}
\newblock Cogam: measuring and moderating cognitive load in machine learning model explanations.
\newblock In {\em Proceedings of the 2020 CHI Conference on Human Factors in Computing Systems\/} (2020), pp.~1--14.

\bibitem{abowd2018us}
{\sc Abowd, J.~M.}
\newblock {The U.S. Census Bureau Adopts Differential Privacy}.
\newblock In {\em Proceedings of the 24th ACM SIGKDD International Conference on Knowledge Discovery \& Data Mining\/} (New York, NY, USA, 2018), KDD '18, Association for Computing Machinery, p.~2867.

\bibitem{abu2018exploring}
{\sc Abu-Salma, R., Redmiles, E.~M., Ur, B., and Wei, M.}
\newblock Exploring user mental models of $\{$End-to-End$\}$ encrypted communication tools.
\newblock In {\em 8th USENIX Workshop on Free and Open Communications on the Internet (FOCI 18)\/} (2018).

\bibitem{alaqracommunicating}
{\sc Alaqra, A.~S., Karegar, F., and Fischer-H{\"u}bner, S.}
\newblock Communicating the privacy functionality of {PETs} to {eHealth} stakeholders.

\bibitem{alaqra2023structural}
{\sc Alaqra, A.~S., Karegar, F., and Fischer-H{\"u}bner, S.}
\newblock Structural and functional explanations for informing lay and expert users: the case of functional encryption.
\newblock {\em Proceedings on Privacy Enhancing Technologies 4\/} (2023), 359--380.

\bibitem{awsnitro}
{\sc {Amazon Web Services}}.
\newblock {Nitro Enclaves}.
\newblock \url{https://aws.amazon.com/ec2/nitro/nitro-enclaves/}.
\newblock Accessed 9/14/2023.

\bibitem{apple2017learning}
{\sc Apple, D. P.~T.}
\newblock Learning with privacy at scale.
\newblock {\em Apple Machine Learning Journal 1}, 8 (2017).

\bibitem{asgharpour2007mental}
{\sc Asgharpour, F., Liu, D., and Camp, L.~J.}
\newblock Mental models of security risks.
\newblock In {\em Financial Cryptography and Data Security\/} (Berlin, Heidelberg, 2007), S.~Dietrich and R.~Dhamija, Eds., Springer Berlin Heidelberg, pp.~367--377.

\bibitem{benthall2022}
{\sc Benthall, S., and Cummings, R.}
\newblock Integrating differential privacy and contextual integrity.
\newblock USENIX Association.

\bibitem{bhaskar2010discovering}
{\sc Bhaskar, R., Laxman, S., Smith, A., and Thakurta, A.}
\newblock Discovering frequent patterns in sensitive data.
\newblock In {\em Proceedings of the 16th ACM SIGKDD international conference on Knowledge discovery and data mining\/} (2010), pp.~503--512.

\bibitem{Bichsel2021}
{\sc Bichsel, B., Steffen, S., Bogunovic, I., and Vechev, M.}
\newblock Dp-sniper: Black-box discovery of differential privacy violations using classifiers.
\newblock In {\em 2021 IEEE Symposium on Security and Privacy (SP)\/} (2021), pp.~391--409.

\bibitem{bravo-lillo-11-warning-mental}
{\sc Bravo-Lillo, C., Cranor, L.~F., Downs, J.~S., and Komanduri, S.}
\newblock {Bridging the Gap in Computer Security Warnings: A Mental Model Approach}.
\newblock {\em IEEE Security \& Privacy 9}, 2 (Mar. 2011), 18--26.

\bibitem{Bullek2017}
{\sc Bullek, B., Garboski, S., Mir, D.~J., and Peck, E.~M.}
\newblock Towards {Understanding} {Differential} {Privacy}: {When} {Do} {People} {Trust} {Randomized} {Response} {Technique}?
\newblock In {\em Proceedings of the 2017 {CHI} {Conference} on {Human} {Factors} in {Computing} {Systems}\/} (Denver, Colorado, USA, May 2017), {CHI} '17, Association for Computing Machinery, pp.~3833--3837.

\bibitem{camp2009mental}
{\sc Camp, L.~J.}
\newblock Mental models of privacy and security.
\newblock {\em IEEE Technology and society magazine 28}, 3 (2009), 37--46.

\bibitem{CCS:CSVW22}
{\sc Casacuberta, S., Shoemate, M., Vadhan, S.~P., and Wagaman, C.}
\newblock Widespread underestimation of sensitivity in differentially private libraries and how to fix it.
\newblock In {\em ACM CCS 2022\/} (Nov. 2022), H.~Yin, A.~Stavrou, C.~Cremers, and E.~Shi, Eds., {ACM} Press, pp.~471--484.

\bibitem{lockicons}
{\sc {Chromium Blog}}.
\newblock An update on the lock icon, May 2023.

\bibitem{craik1967nature}
{\sc Craik, K. J.~W.}
\newblock {\em The nature of explanation}, vol.~445.
\newblock CUP Archive, 1967.

\bibitem{cranor2022mobile}
{\sc Cranor, L.~F.}
\newblock Mobile-app privacy nutrition labels missing key ingredients for success.
\newblock {\em Communications of the ACM 65}, 11 (2022), 26--28.

\bibitem{CCS:CumKapRed21}
{\sc Cummings, R., Kaptchuk, G., and Redmiles, E.~M.}
\newblock ``{I} need a better description'': An investigation into user expectations for differential privacy.
\newblock In {\em ACM CCS 2021\/} (Nov. 2021), G.~Vigna and E.~Shi, Eds., {ACM} Press, pp.~3037--3052.

\bibitem{dankar2013practicing}
{\sc Dankar, F.~K., and El~Emam, K.}
\newblock Practicing differential privacy in health care: A review.
\newblock {\em Trans. Data Priv. 6}, 1 (2013), 35--67.

\bibitem{DCHL23}
{\sc Dekel, I., Cummings, R., Heffetz, O., and Ligett, K.}
\newblock The privacy elasticity of behavior: Conceptualization and application.
\newblock In {\em Proceedings of the 24th ACM Conference on Economics and Computation\/} (2023), EC '23.

\bibitem{demjaha2018metaphors}
{\sc Demjaha, A., Spring, J.~M., Becker, I., Parkin, S., and Sasse, M.~A.}
\newblock Metaphors considered harmful? an exploratory study of the effectiveness of functional metaphors for end-to-end encryption.
\newblock In {\em Proc. USEC\/} (2018), vol.~2018, Internet Society.

\bibitem{desfontainesblog20211001}
{\sc Desfontaines, D.}
\newblock A list of real-world uses of differential privacy.
\newblock \url{https://desfontain.es/privacy/real-world-differential-privacy.html}, 10 2021.
\newblock Ted is writing things (personal blog).

\bibitem{desfontaines2020sok}
{\sc Desfontaines, D., and Pej{\'o}, B.}
\newblock Sok: differential privacies.
\newblock {\em Proceedings on privacy enhancing technologies 2020}, 2 (2020), 288--313.

\bibitem{CCS:DWWZK18}
{\sc Ding, Z., Wang, Y., Wang, G., Zhang, D., and Kifer, D.}
\newblock Detecting violations of differential privacy.
\newblock In {\em ACM CCS 2018\/} (Oct. 2018), D.~Lie, M.~Mannan, M.~Backes, and X.~Wang, Eds., {ACM} Press, pp.~475--489.

\bibitem{distler2020making}
{\sc Distler, V., Lallemand, C., and Koenig, V.}
\newblock Making encryption feel secure: Investigating how descriptions of encryption impact perceived security.
\newblock In {\em 2020 IEEE European Symposium on Security and Privacy Workshops (EuroS\&PW)\/} (2020), IEEE, pp.~220--229.

\bibitem{dwork2019differential}
{\sc Dwork, C., Kohli, N., and Mulligan, D.}
\newblock Differential privacy in practice: Expose your epsilons!
\newblock {\em Journal of Privacy and Confidentiality 9}, 2 (2019).

\bibitem{TCC:DMNS06}
{\sc Dwork, C., McSherry, F., Nissim, K., and Smith, A.}
\newblock Calibrating noise to sensitivity in private data analysis.
\newblock In {\em TCC~2006\/} (Mar. 2006), S.~Halevi and T.~Rabin, Eds., vol.~3876 of {\em {LNCS}}, Springer, Heidelberg, pp.~265--284.

\bibitem{dwork2006calibrating}
{\sc Dwork, C., McSherry, F., Nissim, K., and Smith, A.}
\newblock Calibrating noise to sensitivity in private data analysis.
\newblock In {\em Theory of cryptography conference\/} (2006), Springer, pp.~265--284.

\bibitem{emami2022informative}
{\sc Emami-Naeini, P., Dheenadhayalan, J., Agarwal, Y., and Cranor, L.~F.}
\newblock An informative security and privacy “nutrition” label for internet of things devices.
\newblock {\em IEEE Security \& Privacy 20}, 02 (2022), 31--39.

\bibitem{erlingsson2014rappor}
{\sc Erlingsson, {\'U}., Pihur, V., and Korolova, A.}
\newblock {RAPPOR}: Randomized aggregatable privacy-preserving ordinal response.
\newblock In {\em Proceedings of the 2014 ACM SIGSAC Conference on Computer and Communications Security\/} (2014), pp.~1054--1067.

\bibitem{felt2016rethinking}
{\sc Felt, A.~P., Reeder, R.~W., Ainslie, A., Harris, H., Walker, M., Thompson, C., Acer, M.~E., Morant, E., and Consolvo, S.}
\newblock Rethinking connection security indicators.
\newblock In {\em Twelfth Symposium on Usable Privacy and Security (SOUPS 2016)\/} (2016), pp.~1--14.

\bibitem{10.1145/2556288.2557292}
{\sc Felt, A.~P., Reeder, R.~W., Almuhimedi, H., and Consolvo, S.}
\newblock Experimenting at scale with google chrome's ssl warning.
\newblock In {\em Proceedings of the SIGCHI Conference on Human Factors in Computing Systems\/} (New York, NY, USA, 2014), CHI '14, Association for Computing Machinery, p.~2667–2670.

\bibitem{CCS:FVSTM22}
{\sc Franzen, D., von Voigt, S.~N., S{\"o}rries, P., Tschorsch, F., and M{\"u}ller-Birn, C.}
\newblock Am {I} private and if so, how many?: {C}ommunicating privacy guarantees of differential privacy with risk communication formats.
\newblock In {\em ACM CCS 2022\/} (Nov. 2022), H.~Yin, A.~Stavrou, C.~Cremers, and E.~Shi, Eds., {ACM} Press, pp.~1125--1139.

\bibitem{frik2023model}
{\sc Frik, A., Bernd, J., and Egelman, S.}
\newblock A model of contextual factors affecting older adults’ information-sharing decisions in the us.
\newblock {\em ACM Transactions on Computer-Human Interaction 30}, 1 (2023), 1--48.

\bibitem{gluck2016short}
{\sc Gluck, J., Schaub, F., Friedman, A., Habib, H., Sadeh, N., Cranor, L.~F., and Agarwal, Y.}
\newblock How short is too short? implications of length and framing on the effectiveness of privacy notices.
\newblock In {\em Twelfth symposium on usable privacy and security (SOUPS 2016)\/} (2016), USENIX Association, pp.~321--340.

\bibitem{gluck-16-short-notices}
{\sc Gluck, J., Schaub, F., Friedman, A., Habib, H., Sadeh, N., Cranor, L.~F., and Agarwal, Y.}
\newblock {How Short Is Too Short? Implications of Length and Framing on the Effectiveness of Privacy Notices}.
\newblock In {\em Symposium on Usable Privacy and Security\/} (Denver, Colorado, USA, July 2016), SOUPS~'16, USENIX, pp.~321--340.

\bibitem{golla2018site}
{\sc Golla, M., Wei, M., Hainline, J., Filipe, L., D{\"u}rmuth, M., Redmiles, E., and Ur, B.}
\newblock " what was that site doing with my facebook password?" designing password-reuse notifications.
\newblock In {\em Proceedings of the 2018 ACM SIGSAC Conference on Computer and Communications Security\/} (2018), pp.~1549--1566.

\bibitem{habib2018away}
{\sc Habib, H., Colnago, J., Gopalakrishnan, V., Pearman, S., Thomas, J., Acquisti, A., Christin, N., and Cranor, L.~F.}
\newblock Away from prying eyes: Analyzing usage and understanding of private browsing.
\newblock In {\em Fourteenth symposium on usable privacy and security (SOUPS 2018)\/} (2018), pp.~159--175.

\bibitem{hargittai2012succinct}
{\sc Hargittai, E., and Hsieh, Y.~P.}
\newblock Succinct survey measures of web-use skills.
\newblock {\em Social Science Computer Review 30}, 1 (2012), 95--107.

\bibitem{hargittai2019internet}
{\sc Hargittai, E., and Micheli, M.}
\newblock Internet skills and why they matter.
\newblock {\em Society and the internet: How networks of information and communication are changing our lives 109\/} (2019).

\bibitem{herzberg2016can}
{\sc Herzberg, A., and Leibowitz, H.}
\newblock Can johnny finally encrypt? evaluating e2e-encryption in popular im applications.
\newblock In {\em Proceedings of the 6th Workshop on Socio-Technical Aspects in Security and Trust\/} (2016), pp.~17--28.

\bibitem{holland2020dataset}
{\sc Holland, S., Hosny, A., Newman, S., Joseph, J., and Chmielinski, K.}
\newblock The dataset nutrition label.
\newblock {\em Data Protection and Privacy, Volume 12: Data Protection and Democracy 12\/} (2020), 1.

\bibitem{hullman2015hypothetical}
{\sc Hullman, J., Resnick, P., and Adar, E.}
\newblock Hypothetical outcome plots outperform error bars and violin plots for inferences about reliability of variable ordering.
\newblock {\em PloS one 10}, 11 (2015), e0142444.

\bibitem{ion2011home}
{\sc Ion, I., Sachdeva, N., Kumaraguru, P., and {\v{C}}apkun, S.}
\newblock Home is safer than the cloud! privacy concerns for consumer cloud storage.
\newblock In {\em Proceedings of the Seventh Symposium on Usable Privacy and Security\/} (2011), pp.~1--20.

\bibitem{jin2022}
{\sc Jin, J., McMurtry, E., Rubinstein, B. I.~P., and Ohrimenko, O.}
\newblock Are we there yet? timing and floating-point attacks on differential privacy systems.
\newblock In {\em 2022 IEEE Symposium on Security and Privacy (SP)\/} (2022), pp.~473--488.

\bibitem{johnson2020chorus}
{\sc Johnson, N., Near, J.~P., Hellerstein, J.~M., and Song, D.}
\newblock Chorus: a programming framework for building scalable differential privacy mechanisms.
\newblock In {\em 2020 IEEE European Symposium on Security and Privacy (EuroS\&P)\/} (2020), IEEE, pp.~535--551.

\bibitem{jordon2018pate}
{\sc Jordon, J., Yoon, J., and Van Der~Schaar, M.}
\newblock Pate-gan: Generating synthetic data with differential privacy guarantees.
\newblock In {\em International conference on learning representations\/} (2018).

\bibitem{kacsmar2022comprehension}
{\sc Kacsmar, B., Duddu, V., Tilbury, K., Ur, B., and Kerschbaum, F.}
\newblock Comprehension from chaos: What users understand and expect from private computation.
\newblock {\em arXiv preprint arXiv:2211.07026\/} (2022).

\bibitem{kang2015my}
{\sc Kang, R., Dabbish, L., Fruchter, N., and Kiesler, S.}
\newblock my data just goes everywhere:” user mental models of the internet and implications for privacy and security.
\newblock In {\em Eleventh Symposium on Usable Privacy and Security (SOUPS 2015)\/} (2015), Ottawa, pp.~39--52.

\bibitem{karegar2022exploring}
{\sc Karegar, F., Alaqra, A.~S., and Fischer-H{\"u}bner, S.}
\newblock Exploring $\{$User-Suitable$\}$ metaphors for differentially private data analyses.
\newblock In {\em Eighteenth Symposium on Usable Privacy and Security (SOUPS 2022)\/} (2022), pp.~175--193.

\bibitem{kasiviswanathan2011can}
{\sc Kasiviswanathan, S.~P., Lee, H.~K., Nissim, K., Raskhodnikova, S., and Smith, A.}
\newblock What can we learn privately?
\newblock {\em SIAM Journal on Computing 40}, 3 (2011), 793--826.

\bibitem{kelley2009nutrition}
{\sc Kelley, P.~G., Bresee, J., Cranor, L.~F., and Reeder, R.~W.}
\newblock A" nutrition label" for privacy.
\newblock In {\em Proceedings of the 5th Symposium on Usable Privacy and Security\/} (2009), pp.~1--12.

\bibitem{kelley2010standardizing}
{\sc Kelley, P.~G., Cesca, L., Bresee, J., and Cranor, L.~F.}
\newblock Standardizing privacy notices: an online study of the nutrition label approach.
\newblock In {\em Proceedings of the SIGCHI Conference on Human factors in Computing Systems\/} (2010), pp.~1573--1582.

\bibitem{kifer2011no}
{\sc Kifer, D., and Machanavajjhala, A.}
\newblock No free lunch in data privacy.
\newblock In {\em Proceedings of the 2011 ACM SIGMOD International Conference on Management of data\/} (2011), pp.~193--204.

\bibitem{kifer2014pufferfish}
{\sc Kifer, D., and Machanavajjhala, A.}
\newblock Pufferfish: A framework for mathematical privacy definitions.
\newblock {\em ACM Transactions on Database Systems (TODS) 39}, 1 (2014), 1--36.

\bibitem{kifer2020guidelines}
{\sc Kifer, D., Messing, S., Roth, A., Thakurta, A., and Zhang, D.}
\newblock Guidelines for implementing and auditing differentially private systems.
\newblock {\em arXiv preprint arXiv:2002.04049\/} (2020).

\bibitem{kollnig2022goodbye}
{\sc Kollnig, K., Shuba, A., Van~Kleek, M., Binns, R., and Shadbolt, N.}
\newblock Goodbye tracking? impact of ios app tracking transparency and privacy labels.
\newblock In {\em 2022 ACM Conference on Fairness, Accountability, and Transparency\/} (2022), pp.~508--520.

\bibitem{krombholz2019if}
{\sc Krombholz, K., Busse, K., Pfeffer, K., Smith, M., and Von~Zezschwitz, E.}
\newblock " if https were secure, i wouldn't need 2fa"-end user and administrator mental models of https.
\newblock In {\em 2019 IEEE Symposium on Security and Privacy (SP)\/} (2019), IEEE, pp.~246--263.

\bibitem{kuhtreiber2022replication}
{\sc K{\"u}htreiber, P., Pak, V., and Reinhardt, D.}
\newblock Replication: The effect of differential privacy communication on german users' comprehension and data sharing attitudes.
\newblock In {\em Eighteenth Symposium on Usable Privacy and Security (SOUPS 2022)\/} (2022), pp.~117--134.

\bibitem{10.1145/3209811.3212701}
{\sc Lapets, A., Jansen, F., Albab, K.~D., Issa, R., Qin, L., Varia, M., and Bestavros, A.}
\newblock Accessible privacy-preserving web-based data analysis for assessing and addressing economic inequalities.
\newblock In {\em Proceedings of the 1st ACM SIGCAS Conference on Computing and Sustainable Societies\/} (New York, NY, USA, 2018), COMPASS '18, Association for Computing Machinery.

\bibitem{li2022understanding}
{\sc Li, T., Reiman, K., Agarwal, Y., Cranor, L.~F., and Hong, J.~I.}
\newblock Understanding challenges for developers to create accurate privacy nutrition labels.
\newblock In {\em Proceedings of the 2022 CHI Conference on Human Factors in Computing Systems\/} (2022), pp.~1--24.

\bibitem{lin2012expectation}
{\sc Lin, J., Amini, S., Hong, J.~I., Sadeh, N., Lindqvist, J., and Zhang, J.}
\newblock Expectation and purpose: understanding users' mental models of mobile app privacy through crowdsourcing.
\newblock In {\em Proceedings of the 2012 ACM conference on ubiquitous computing\/} (2012), pp.~501--510.

\bibitem{lipford2009visible}
{\sc Lipford, H.~R., Hull, G., Latulipe, C., Besmer, A., and Watson, J.}
\newblock Visible flows: Contextual integrity and the design of privacy mechanisms on social network sites.
\newblock In {\em 2009 International Conference on Computational Science and Engineering\/} (2009), vol.~4, IEEE, pp.~985--989.

\bibitem{lyu2017}
{\sc Lyu, M., Su, D., and Li, N.}
\newblock Understanding the sparse vector technique for differential privacy.
\newblock {\em Proc. VLDB Endow. 10}, 6 (feb 2017), 637–648.

\bibitem{ma2019impact}
{\sc Ma, Z., Reynolds, J., Dickinson, J., Wang, K., Judd, T., Barnes, J.~D., Mason, J., and Bailey, M.}
\newblock The impact of secure transport protocols on phishing efficacy.
\newblock In {\em 12th USENIX Workshop on Cyber Security Experimentation and Test (CSET 19)\/} (2019).

\bibitem{mcdonald2008cost}
{\sc McDonald, A.~M., and Cranor, L.~F.}
\newblock The cost of reading privacy policies.
\newblock {\em Isjlp 4\/} (2008), 543.

\bibitem{McDonald2019}
{\sc McDonald, N., Schoenebeck, S., and Forte, A.}
\newblock Reliability and inter-rater reliability in qualitative research: Norms and guidelines for cscw and hci practice.
\newblock {\em Proc. ACM Hum.-Comput. Interact. 3}, CSCW (nov 2019).

\bibitem{mcsherry2009}
{\sc McSherry, F.~D.}
\newblock Privacy integrated queries: An extensible platform for privacy-preserving data analysis.
\newblock In {\em Proceedings of the 2009 ACM SIGMOD International Conference on Management of Data\/} (New York, NY, USA, 2009), SIGMOD '09, Association for Computing Machinery, p.~19–30.

\bibitem{fb2020}
{\sc Messing, S., DeGregorio, C., Hillenbrand, B., King, G., Mahanti, S., Mukerjee, Z., Nayak, C., Persily, N., State, B., and Wilkins, A.}
\newblock {Facebook Privacy-Protected Full URLs Data Set}, 2020.

\bibitem{sealfhe}
{\sc {Microsoft}}.
\newblock {Microsoft SEAL: Fast and Easy-to-Use Homomorphic Encryption Library}.
\newblock \url{https://www.microsoft.com/en-us/research/project/microsoft-seal/}.
\newblock Accessed 9/14/2023.

\bibitem{tumult}
{\sc Miklau, G.}
\newblock {How Tumult Labs helped the IRS support educational accountability with differential privacy}, July 2021.

\bibitem{CCS:Mironov12}
{\sc Mironov, I.}
\newblock On significance of the least significant bits for differential privacy.
\newblock In {\em ACM CCS 2012\/} (Oct. 2012), T.~Yu, G.~Danezis, and V.~D. Gligor, Eds., {ACM} Press, pp.~650--661.

\bibitem{naeini2017privacy}
{\sc Naeini, P.~E., Bhagavatula, S., Habib, H., Degeling, M., Bauer, L., Cranor, L.~F., and Sadeh, N.}
\newblock Privacy expectations and preferences in an iot world.
\newblock In {\em Thirteenth Symposium on Usable Privacy and Security (SOUPS 2017)\/} (2017), USENIX Association Santa Clara, pp.~399--412.

\bibitem{nanayakkara2022visualizing}
{\sc Nanayakkara, P., Bater, J., He, X., Hullman, J.~R., and Duggan, J.}
\newblock Visualizing privacy-utility trade-offs in differentially private data releases.
\newblock {\em Proceedings on Privacy Enhancing Technologies 2022\/} (2022), 601 -- 618.

\bibitem{nanayakkara2022s}
{\sc Nanayakkara, P., and Hullman, J.}
\newblock What's driving conflicts around differential privacy for the us census.
\newblock {\em IEEE Security \& Privacy}, 01 (2022), 2--11.

\bibitem{nanayakkara2023chances}
{\sc Nanayakkara, P., Smart, M.~A., Cummings, R., Kaptchuk, G., and Redmiles, E.}
\newblock What are the chances? explaining the epsilon parameter in differential privacy.
\newblock {\em arXiv preprint arXiv:2303.00738\/} (2023).

\bibitem{nist}
{\sc Near, J., and Darais, D.}
\newblock { Threat Models for Differential Privacy }, 2020.

\bibitem{nissenbaum2009privacy}
{\sc Nissenbaum, H.}
\newblock Privacy in context.
\newblock In {\em Privacy in Context}. Stanford University Press, 2009.

\bibitem{oates2018turtles}
{\sc Oates, M., Ahmadullah, Y., Marsh, A., Swoopes, C., Zhang, S., Balebako, R., and Cranor, L.~F.}
\newblock Turtles, locks, and bathrooms: Understanding mental models of privacy through illustration.
\newblock {\em Proceedings on Privacy Enhancing Technologies 2018}, 4 (2018), 5--32.

\bibitem{Obar2020}
{\sc Obar, J.~A., and Oeldorf-Hirsch, A.}
\newblock The biggest lie on the internet: Ignoring the privacy policies and terms of service policies of social networking services.
\newblock {\em Information, Communication \& Society 23}, 1 (2020), 128--147.

\bibitem{pair}
{\sc Pearce, A., and Jiang, E.}
\newblock {How randomized response can help collect sensitive information responsibly}, 2020.

\bibitem{pew_mobile}
{\sc {Pew Research Center}}.
\newblock Mobile fact sheet.
\newblock \url{https://www.pewresearch.org/internet/fact-sheet/mobile/}, Apr 2021.
\newblock Accessed 7/29/2022.

\bibitem{raja2011brick}
{\sc Raja, F., Hawkey, K., Hsu, S., Wang, K.-L.~C., and Beznosov, K.}
\newblock A brick wall, a locked door, and a bandit: a physical security metaphor for firewall warnings.
\newblock In {\em Proceedings of the seventh symposium on usable privacy and security\/} (2011), pp.~1--20.

\bibitem{rao-16-unexpected}
{\sc Rao, A., Schaub, F., Sadeh, N., Acquisti, A., and Kang, R.}
\newblock {Expecting the Unexpected: Understanding Mismatched Privacy Expectations Online}.
\newblock In {\em Symposium on Usable Privacy and Security\/} (Denver, Colorado, USA, July 2016), SOUPS~'16, USENIX, pp.~77--96.

\bibitem{redmiles2017you}
{\sc Redmiles, E.~M., Liu, E., and Mazurek, M.~L.}
\newblock You want me to do what? a design study of two-factor authentication messages.
\newblock In {\em SOUPS\/} (2017), vol.~57, p.~93.

\bibitem{CCS:RZKKDM18}
{\sc Redmiles, E.~M., Zhu, Z., Kross, S., Kuchhal, D., Dumitras, T., and Mazurek, M.~L.}
\newblock Asking for a friend: Evaluating response biases in security user studies.
\newblock In {\em ACM CCS 2018\/} (Oct. 2018), D.~Lie, M.~Mannan, M.~Backes, and X.~Wang, Eds., {ACM} Press, pp.~1238--1255.

\bibitem{reed2010}
{\sc Reed, J., and Pierce, B.~C.}
\newblock Distance makes the types grow stronger: A calculus for differential privacy.
\newblock {\em SIGPLAN Not. 45}, 9 (sep 2010), 157–168.

\bibitem{rogaway2015moral}
{\sc Rogaway, P.}
\newblock The moral character of cryptographic work.
\newblock {\em Cryptology ePrint Archive\/} (2015).

\bibitem{schaub-15-effective-notices}
{\sc Schaub, F., Balebako, R., Durity, A.~L., and Cranor, L.~F.}
\newblock {A Design Space for Effective Privacy Notices}.
\newblock In {\em Symposium on Usable Privacy and Security\/} (Ottawa, Canada, July 2015), SOUPS~'15, USENIX, pp.~1--17.

\bibitem{schomakers2019internet}
{\sc Schomakers, E.-M., Lidynia, C., M{\"u}llmann, D., and Ziefle, M.}
\newblock Internet users’ perceptions of information sensitivity--insights from germany.
\newblock {\em International Journal of Information Management 46\/} (2019), 142--150.

\bibitem{smart2022}
{\sc Smart, M.~A., Sood, D., and Vaccaro, K.}
\newblock Understanding risks of privacy theater with differential privacy.
\newblock {\em Proc. ACM Hum.-Comput. Interact. 6}, CSCW2 (nov 2022).

\bibitem{stewart2012death}
{\sc Stewart, G., and Lacey, D.}
\newblock Death by a thousand facts: Criticising the technocratic approach to information security awareness.
\newblock {\em Information Management \& Computer Security 20}, 1 (2012), 29--38.

\bibitem{stransky2021limited}
{\sc Stransky, C., Wermke, D., Schrader, J., Huaman, N., Acar, Y., Fehlhaber, A.~L., Wei, M., Ur, B., and Fahl, S.}
\newblock On the limited impact of visualizing encryption: Perceptions of e2e messaging security.
\newblock In {\em Seventeenth Symposium on Usable Privacy and Security\/} (2021), pp.~437--454.

\bibitem{suh2022privacytoon}
{\sc Suh, S., Lamorea, S., Law, E., and Zhang-Kennedy, L.}
\newblock Privacytoon: Concept-driven storytelling with creativity support for privacy concepts.
\newblock In {\em Designing Interactive Systems Conference\/} (2022), pp.~41--57.

\bibitem{tang2022well}
{\sc Tang, J., Birrell, E., and Lerner, A.}
\newblock How well do my results generalize now? the external validity of online privacy and security surveys.
\newblock {\em arXiv preprint arXiv:2202.14036\/} (2022).

\bibitem{thakurta2017learning}
{\sc Thakurta, A.~G., Vyrros, A.~H., Vaishampayan, U.~S., Kapoor, G., Freudiger, J., Sridhar, V.~R., and Davidson, D.}
\newblock Learning new words, Mar.~14 2017.
\newblock US Patent 9,594,741.

\bibitem{tramer2022debugging}
{\sc Tramer, F., Terzis, A., Steinke, T., Song, S., Jagielski, M., and Carlini, N.}
\newblock Debugging differential privacy: A case study for privacy auditing.
\newblock {\em arXiv preprint arXiv:2202.12219\/} (2022).

\bibitem{Turow2018}
{\sc Turow, J., Hennessy, M., and Draper, N.}
\newblock Persistent misperceptions: Americans’ misplaced confidence in privacy policies, 2003--2015.
\newblock {\em Journal of Broadcasting \& Electronic Media 62}, 3 (2018), 461--478.

\bibitem{vaniea2014mental}
{\sc Vaniea, K., Rader, E., and Wash, R.}
\newblock Mental models of software updates.
\newblock {\em International Communication Association\/} (2014), 1--39.

\bibitem{Warner1965}
{\sc Warner, S.~L.}
\newblock Randomized {Response}: {A} {Survey} {Technique} for {Eliminating} {Evasive} {Answer} {Bias}.
\newblock {\em Journal of the American Statistical Association 60}, 309 (Mar. 1965), 63--69.

\bibitem{wash2010folk}
{\sc Wash, R.}
\newblock Folk models of home computer security.
\newblock In {\em Proceedings of the Sixth Symposium on Usable Privacy and Security\/} (2010), pp.~1--16.

\bibitem{whitten1999johnny}
{\sc Whitten, A., and Tygar, J.~D.}
\newblock Why johnny can't encrypt: A usability evaluation of pgp 5.0.
\newblock In {\em USENIX security symposium\/} (1999), vol.~348, pp.~169--184.

\bibitem{xiong2020effect}
{\sc Xiong, A.}
\newblock Effect of facts box on users’ comprehension of differential privacy: A preliminary study.
\newblock In {\em Proceedings of the Human Factors and Ergonomics Society 2020 Annual Meeting\/} (2020).

\bibitem{xiong2020towards}
{\sc Xiong, A., Wang, T., Li, N., and Jha, S.}
\newblock Towards effective differential privacy communication for users’ data sharing decision and comprehension.
\newblock In {\em 2020 IEEE Symposium on Security and Privacy (SP)\/} (2020), IEEE, pp.~392--410.

\bibitem{Xiong2022UsingIT}
{\sc Xiong, A., Wu, C., Wang, T., Proctor, R.~W., Blocki, J., Li, N., and Jha, S.}
\newblock Using illustrations to communicate differential privacy trust models: An investigation of users' comprehension, perception, and data sharing decision.
\newblock {\em ArXiv abs/2202.10014\/} (2022).

\bibitem{zhang2017}
{\sc Zhang, D., and Kifer, D.}
\newblock Lightdp: Towards automating differential privacy proofs.
\newblock In {\em Proceedings of the 44th ACM SIGPLAN Symposium on Principles of Programming Languages\/} (New York, NY, USA, 2017), POPL '17, Association for Computing Machinery, p.~888–901.

\bibitem{ZhangKennedy}
{\sc Zhang-Kennedy, L., Chiasson, S., and Biddle, R.}
\newblock Password advice shouldn't be boring: Visualizing password guessing attacks.
\newblock In {\em 2013 APWG eCrime Researchers Summit\/} (2013), pp.~1--11.

\bibitem{zhang2016role}
{\sc Zhang-Kennedy, L., Chiasson, S., and Biddle, R.}
\newblock The role of instructional design in persuasion: A comics approach for improving cybersecurity.
\newblock {\em International Journal of Human-Computer Interaction 32}, 3 (2016), 215--257.

\end{thebibliography}
